\documentclass[fleqn,usenatbib]{mnras}

\usepackage{newtxtext,newtxmath}
\usepackage[T1]{fontenc}
\usepackage{ae,aecompl}

\usepackage{graphicx}	% Including figure files
\usepackage{amsmath}	% Advanced maths commands
\usepackage{amssymb}	% Extra maths symbols
\usepackage{pdflscape}
\usepackage{afterpage}
\usepackage{rotating}

\newcommand{\chandra}{\textit{Chandra}}
\newcommand{\nustar}{\textit{NuSTAR}}
\newcommand{\suzaku}{{\it Suzaku}}

\newcommand{\xmm}{{\it XMM-Newton}}

\newcommand{\red}{\textcolor{black}}%{}%
\title[High Density Reflection II.]{High Density Reflection Spectroscopy: II. the Density of the Inner Black Hole Accretion Disc in AGN}

\author[J. Jiang et al.]{Jiachen Jiang,$^{1,2}$\thanks{E-mail: jj447@cam.ac.uk} Andrew C. Fabian,$^{1}$ Thomas Dauser,$^{3}$ Luigi Gallo,$^{4}$
\newauthor Javier A. Garc{\'{\i}}a,$^{3,5}$ Erin Kara,$^{6}$ Michael L. Parker,$^{7}$ John A. Tomsick,$^{8}$
\newauthor  Dominic J. Walton$^{1}$ and Christopher S. Reynolds$^{1}$
\\
$^{1}$Institute of Astronomy, University of Cambridge, Madingley Road, Cambridge CB3 0HA, UK\\
$^{2}$Tsinghua Center for Astrophysics, Tsinghua University, Beijing 100084, China\\
$^{3}$Dr. Karl Remeis-Observatory and Erlangen Centre for Astroparticle Physics, Sternwartstr. 7, D-96049 Bamberg, Germany\\
$^{4}$Department of Astronomy and Physics, Saint Mary's University, 923 Robie Street, Halifax, NS, B3H 3C3, Canada \\
$^{5}$Cahill Center for Astronomy and Astrophysics, California Institute of Technology, Pasadena, CA 91125, USA\\
$^{6}$Department of Astronomy, University of Maryland, College Park, MD 20742-2421, USA\\
$^{7}$European Space Agency (ESA), European Space Astronomy Centre (ESAC), E-28691 Villanueva de la Ca\~nada, Spain\\
$^{8}$Space Sciences Laboratory, 7 Gauss Way, University of California, Berkeley, CA 94720-7450, USA\\
}

\date{Accepted XXX. Received YYY; in original form ZZZ}

\pubyear{2019}

\begin{document}
\label{firstpage}
\pagerange{\pageref{firstpage}--\pageref{lastpage}}
\maketitle

\begin{abstract}
We present a high density disc reflection spectral analysis of a sample of 17 Seyfert 1 galaxies to study the inner disc densities at different black hole mass scales and accretion rates. All the available \xmm\ observations in the archive are used. OM observations in the optical/UV band are used to estimate their accretion rates. We find that 65\% of sources in our sample show a disc density significantly higher than $n_{\rm e}=10^{15}$\,cm$^{-3}$, which was assumed in previous reflection-based spectral analyses. The best-fit disc densities show an anti-correlation with black hole mass and mass accretion rate. High density disc reflection model can successfully explain the soft excess emission and significantly reduce inferred iron abundances. We also compare our black hole spin and disc inclination angle measurements with previous analyses.
\end{abstract}


%%%%%%%%%%%%%%%%%%%%%%%%%%%%%%%%%%%%%%%%%%%%%%%%%%

%%%%%%%%%%%%%%%%% BODY OF PAPER %%%%%%%%%%%%%%%%%%

\section{Introduction} \label{intro}

High density disc reflection spectroscopy allows the density of the inner disc to be a free parameter during the analysis of the disc reflection spectra of accreting black hole (BH) systems \footnote{We refer interested readers to the first paper of this series \citep{jiang19} for more introduction of disc reflection spectra of accreting BHs.}. Previously, a fixed disc density of $n_{\rm e}=10^{15}$\,cm$^{-3}$ was commonly assumed. However the density of the inner disc is expected to be higher than this value in the less massive supermassive BHs (SMBHs, $M_{\rm BH}/M_{\odot}=m_{\rm BH}\approx10^{6}$) and stellar-mass BHs ($m_{\rm BH}=4-20$), as predicted by the standard \citet{shakura73} disc model where all the disc energy is radiated away locally. It has been suggested that a high fraction of disc energy has to be dissipated in the corona instead of being radiated away at the disc surface in order to account for the typical spectral energy distribution of a Seyfert galaxy \citep{haardt91}. Theoretically, one will then obtain an even denser disc than the standard \citet{shakura73} model after considering fraction of disc energy that is transferred to the corona \citep{svensson94}. There have also been increasing observational indications for disc densities higher than $10^{15}$\,cm$^{-3}$ in both BH X-ray binaries \citep[XRBs, e.g.][]{tomsick18, jiang19} and active galactic nuclei \citep[AGN, e.g.][]{jiang18, garcia18}. 

%\footnote{The value of $n_{\rm e}$ is reported in the unit of cm$^{-3}$ hereafter.}

\begin{figure*}
    \centering
    \includegraphics[width=16cm]{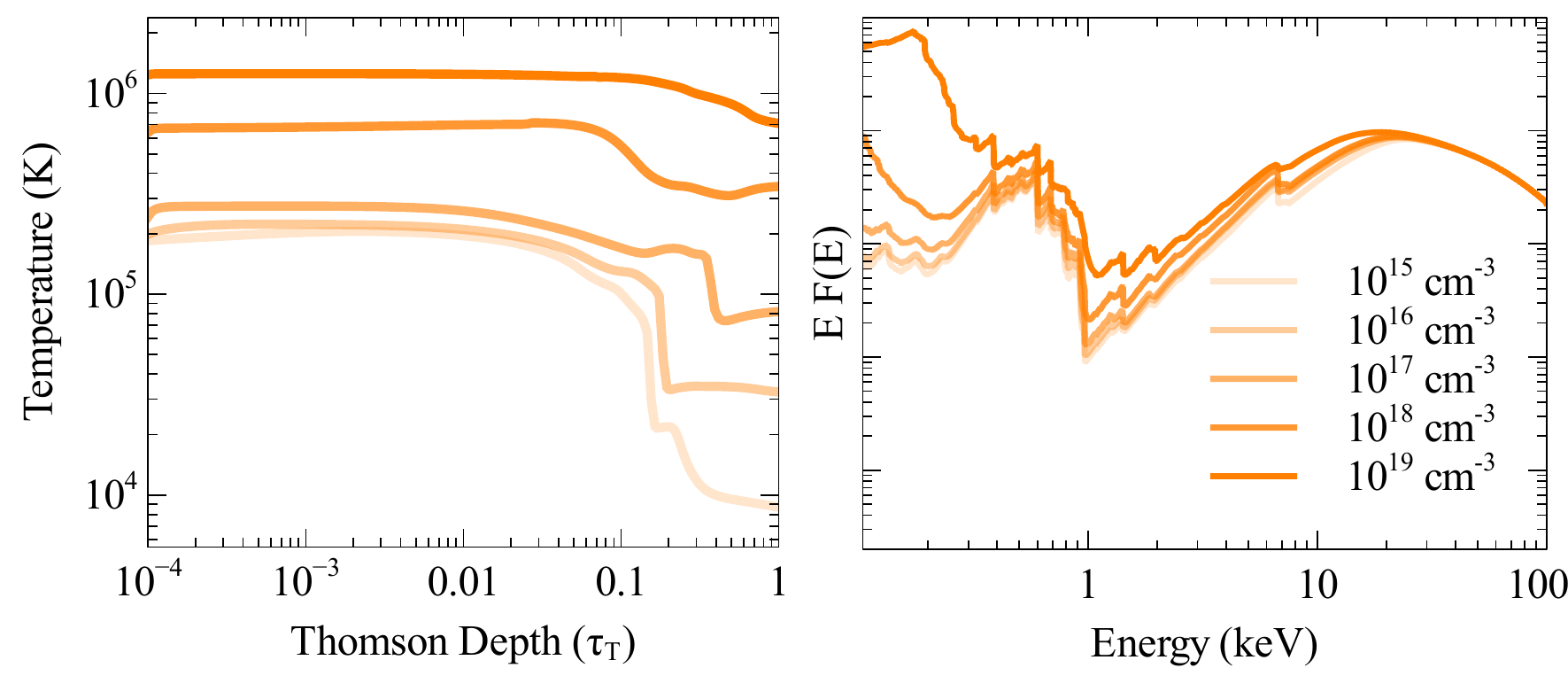}
    \caption{Left: Temperature profiles of an illuminated slab under hydro-static equilibrium for different electron densities calculated by \citet{garcia16}. Right: Relativistic disc reflection spectra calculated by \texttt{relxilld} for different densities corresponding to the left panel. A disc ionization parameter of $\xi=50$, emissivity index of $q=3$, maximum BH spin, and viewing angle of $i=30^{\circ}$ are assumed.}
    \label{pic_ne_spectra}
\end{figure*}

Most reflection simulations calculate spectra from a constant density slab parallel to a thin disc \citep[e.g.][]{ross93,garcia10}. One side of the slab is illuminated and heated up by coronal emission. These simulations then calculate the reflected spectra from the slab by considering both absorption and emission processes. The left panel of Fig.\,\ref{pic_ne_spectra} shows the temperature profiles of an illuminated ionised slab under hydro-static equilibrium \citep{garcia16}. At $n_{\rm e}=10^{15}$\,cm$^{-3}$ and $\xi=50$\,erg\,cm\,s$^{-1}$, the surface temperature of the ionised slab is about $2\times10^{5}$\,K. The temperature remains constant at larger depths before it drops to a lower temperature at Thomson depth $\tau_{\rm T}=0.1$. The drop in the temperature profile is because the recombination (cooling) process dominates over the heating process at $\tau_{\rm T}>0.1$. 

The electron density of the slab changes the temperature profile and thus affects the resulting reflection spectrum. For example, the temperature within $0.1\tau_{\rm T}$ increases by a factor of 10 when $\log(n_{\rm e})$ increases \footnote{The value of $n_{\rm e}$ is reported in the unit of cm$^{-3}$ hereafter.} from 15 to 19. The temperature at $\tau_{\rm T}>0.1$ increases by an even larger factor. See the left panel of Fig.\,\ref{pic_ne_spectra} for temperature profiles for different electron densities. The change of temperature with density is because low energy photons in the disc are trapped by free-free absorption and increases the temperature of the slab at a high electron density \citep{ross07,garcia16}. Consequently, the reflected emission in the soft X-ray band turns into a quasi-blackbody spectrum. See the right panel of Fig.\,\ref{pic_ne_spectra} for corresponding spectra for different disc densities. 

In the first paper of this series, we demonstrate the effects of a high density model on the spectrum of the BH XRB GX~339$-$4. The X-ray spectra of Seyfert galaxies and BH XRBs share many similarities, such as broad Fe~K emission line and Compton hump above 10\,keV \citep[e.g.][]{walton12}. Additionally, the soft X-ray band of Seyfert galaxies is known to show variable excess emission \citep[`soft excess', e.g.][]{arnaud85}. The blackbody-shaped feature in a high density model may be able to explain this soft excess emission \citep[e.g.][]{mallick18,garcia18}. The strongest supporting evidence for the reflection interpretation of the soft excess is the X-ray reverberation lags that have been found in various sources \citep[e.g.][]{fabian09, zoghbi12, demarco13, kara16}.

In this paper, we fit high density models to the \xmm\ spectra of a sample of Seyfert 1 galaxies. By conducting photometry with \xmm\ OM observations, we are able to estimate the mass accretion rate for each BH, and thus compare disc density with BH mass and accretion rate. In Section\,\ref{selection}, we introduce our source sample and data reduction; in Section\,\ref{analysis}, we introduce the analysis method for EPIC-pn and OM observations; in Section\,\ref{results} and \ref{discuss}, we briefly summarise and discuss our results in this work and the first paper of this series. In Appendix\,\ref{re1}, we present a short introduction and details of the EPIC-pn spectral analysis for each individual source. In Appendix\,\ref{gas}, we present an estimation of the disc density at an optical depth of $\tau<1$ of a gas pressure-dominated disc in order to explain the disc densities in previous analyses of BH XRBs. High density disc reflection modelling for the extremely variable narrow-line Seyfert 1 galaxies 1H0707$-$495 will be presented in a future paper of this series.

\section{Sample and Data Reduction} \label{selection}

\begin{table*}
\caption{The sample of Seyfert 1 galaxies considered in this work. The Galactic column density and extinction values are as calculated by \citet{willingale13}. The source distances are from NED. The sixth column shows the total EPIC-pn exposure after filtering flaring particle background. $L_{\rm X}$ is the averaged Galactic absorption-corrected luminosity in the 0.5--10~keV band. The upper table shows all the optical reverberation mapped Seyfert 1 galaxies in the AGN Black Hole Mass Database \citep{bentz15}. The lower table shows a sample of Seyfert 1 galaxies with BH mass measurements obtained by measuring H$\beta$ line width \citep{bianchi09}. $^1$UGC~6728 only has MOS observations as its pn observation is dominated by flaring background.}
\label{tab_bh_info}
\centering
\begin{tabular}{ccccccccc}
\hline\hline
Name & $\log(M_{\rm BH}/M_{\odot})$ & $N_{\rm H}$ ($10^{20}$\,cm$^{-2}$) & $E(B-V)$ & $D$ (Mpc) & Expo (ks) & $\log(L_{\rm X})$ \\
\hline
Ark~120 & $8.07^{+0.05}_{-0.06}$ & 14.5 & 0.126 & 148 & 504& $44.338\pm0.002$\\
Mrk~110 & $7.29\pm0.10$ & 1.39 & 0.014 & 151 & 33 & $44.161^{+0.009}_{-0.006}$\\
Mrk~1310 & $6.21^{+0.07}_{-0.09}$ & 2.66 & 0.029 & 86.7 & 35 & $41.67^{+0.05}_{-0.03}$\\
Mrk~279 & $7.44^{+0.10}_{-0.13}$ & 1.72 & 0.018 & 129 & 113& $43.960^{+0.008}_{-0.124}$\\
Mrk~335 & $7.23\pm0.04$ & 4.11 & 0.044 & 111 & 298 & $43.51^{+0.03}_{-0.02}$\\
Mrk~590 & $7.57^{+0.06}_{-0.07}$  & 2.93 & 0.031 & 107 & 36 & $43.143\pm0.012$\\
Mrk~79  & $7.61^{+0.11}_{-0.14}$  & 6.73 & 0.071 & 94.3 & 100& $43.384^{+0.012}_{-0.013}$\\
NGC~4748 & $6.41^{+0.11}_{-0.18}$ & 4.07 & 0.052 & 65.5 & 26& $43.033\pm0.011$\\
PG~0804$+$761 & $8.74\pm0.05$  & 3.51 & 0.032 & 475 & 32& $44.728\pm0.009$\\
PG~0844$+$349 & $7.86^{+0.15}_{-0.23}$  & 3.13 & 0.032 & 279 & 18& $43.663^{+0.008}_{-0.007}$\\
PG~1229$+$204 & $7.76^{+0.18}_{-0.22}$ & 2.92 & 0.030 & 276 & 17 & $43.80^{+0.03}_{-0.04}$\\
PG~1426$+$015 & $9.01^{+0.11}_{-0.16}$ & 2.88 & 0.033 & 383 & 0.6 & $44.49^{+0.03}_{-0.04}$\\
UGC~6728 & $5.85^{+0.19}_{-0.36}$ & 6.74 & 0.068 & 29.3 & 8.4, 8.8$^{1}$ & $42.053^{+0.015}_{-0.019}$\\
\hline
1H1934$-$063 & $6.5\pm0.5$ & 19.5 & 0.278 & 42.5 & 105& $43.037\pm0.006$\\
Ark~564 & $6.2\pm0.5$ & 6.74 & 0.068 &106 &402& $43.972\pm0.006$\\
Swift~J2127.4$+$5654 & $7.2\pm0.5$ & 91.4 & 1.532 & 61.1 & 326 & $43.380^{+0.005}_{-0.009}$\\
Ton~S180 & $6.9\pm0.5$ & 1.54 & 0.016 & 263 & 148 & $44.01^{+0.07}_{-0.09}$\\
\hline\hline
\end{tabular}
\end{table*}

We select 17 Seyfert 1 galaxies for our work. They include all 13 Seyfert 1 galaxies in the AGN Black Hole Mass Database \citep{bentz15} that have been observed by \textit{XMM-Newton} and 4 Seyfert 1 galaxies where broad Fe~K$\alpha$ emission line was found and supersolar iron abundance was required in previous spectral analyses. The Seyfert 1 galaxies in our sample show no multiple warm absorbers or heavy obscuration along the line of sight. Therefore, we are able to have a clear view of the soft X-ray emission from the innermost regions of their AGN. We ignore the Seyfert 2 galaxies in the AGN Black Hole Mass Database as they are mostly obscured in the X-ray band \citep{risaliti99}. Table\,\ref{tab_bh_info} shows the basic information for each source.

\begin{figure*}
\centering
\includegraphics[width=16cm]{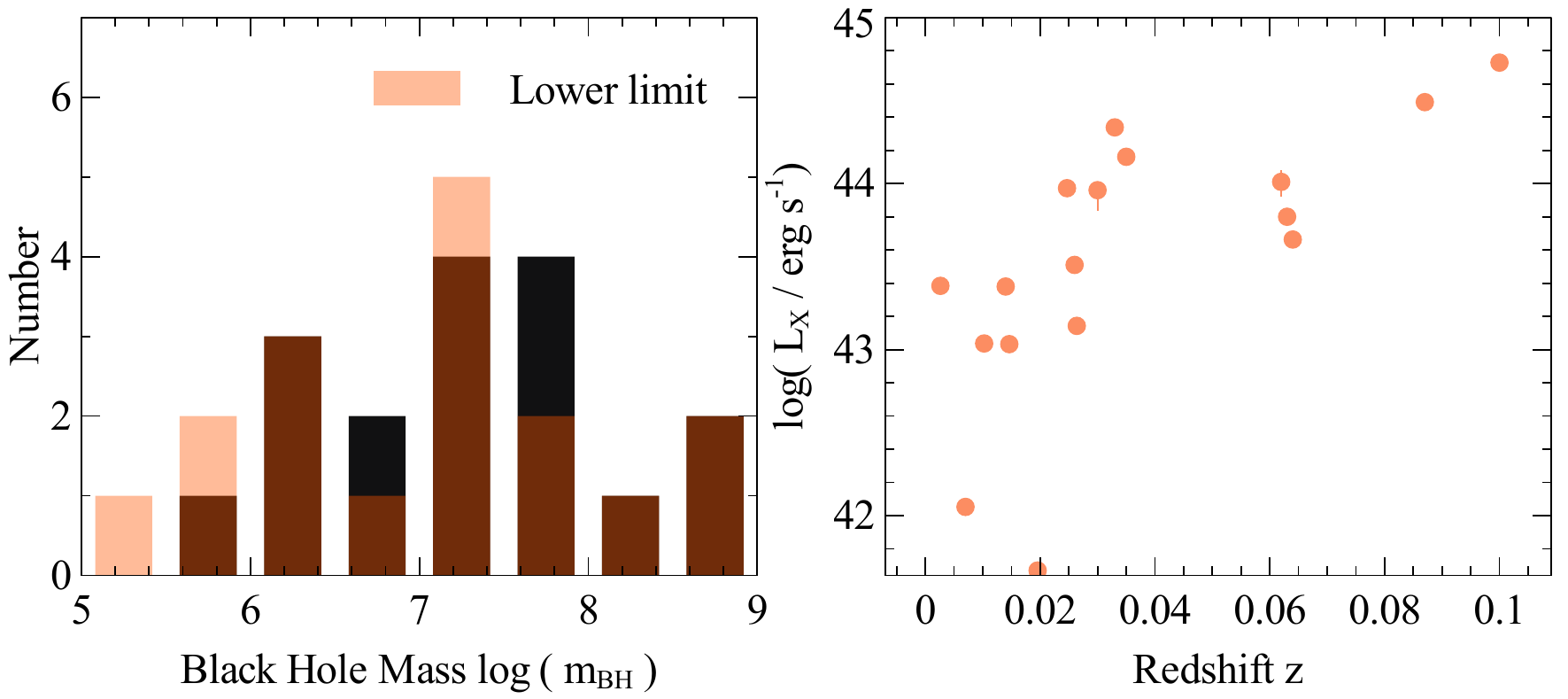}
\caption{Left: black hole mass distribution in our sample. Orange bars show the distribution of the lower limits of mass measurements. Brown shows the overlaps of orange and black bars. Right: source redshift and averaged X-ray luminosity distribution. The luminosities are calculated in the \xmm\ band (0.5--10~keV) after correcting for the Galactic absorption.}
\label{pic_mass}
\end{figure*}

For BHs without a reverberation-mapping mass in the AGN Black Hole Mass Database \citep{bentz15}, we quote the mass measurement from H${\beta}$ line width \citep{bianchi09}. Due to the unknown geometry of the broad line region (BLR), the relation between H$\beta$ line width and BLR velocity distribution has a large uncertainty. For example, \citet{kaspi00} assume a spherical BLR with an isotropic velocity distribution and predict $v_{\rm BLR}=\frac{\sqrt{3}}{2}FWHM(\rm{H}\beta)$. But \citet{mclure04} assume a disc-shaped BLR which predicts a velocity $\sqrt{3}$ times larger and a black hole mass 3 times larger than the values given by \citet{kaspi00}. For this reason, we consider a conservative uncertainty of $\Delta\log(m_{\rm BH})=\pm\log(3)\approx\pm0.5$. The left panel of Fig.\,\ref{pic_mass} shows the BH mass distribution in our sample. 75\% of sources in the sample have a BH mass  between $\log(m_{\rm BH})=6-8$.

We consider all the available archival \xmm\ observations for each AGN. A complete list of the \xmm\ observations considered in this work is in Table\,\ref{tab_obs_info}. We extract EPIC-pn products using SAS 16.1.0, after filtering intervals dominated by flaring particle background. The task EPATPLOT is used to test for any pile-up effects. An annulus-shaped source region is used to extract source products if there is evidence for pile-up. The inner radius of the annulus is chosen to keep distribution patterns consistent with the model curves given by EPATPLOT. Background products are extracted from a nearby region on the same chip, avoiding the areas dominated by background Cu~K emission lines from the underlying electronics. We concentrate on the EPIC-pn spectra between 0.5--10~keV due to its higher effective area compared to the two EPIC-MOS instruments. For UGC~6728, we extract EPIC-MOS spectra as its pn observation is dominated by flaring background. All the spectra are grouped to have a minimum signal-to-noise of 6 and oversample by a factor of 3. Net pn exposures and averaged X-ray luminosities after correcting for Galactic absorption are shown in the last column of Table\,\ref{tab_bh_info}. We stacked spectra from different observations for each source using ADDSPEC. In the right panel of Fig\,\ref{pic_mass}, we present the distribution of the averaged X-ray luminosities in the full pn band and source redshifts of our sample. 

\xmm\ Optical Monitor (OM) data are extracted using the task OMICHAIN, and the count rates are converted to Galactic extinction-corrected flux for each filter. The flux conversion factors are provided by SAS Watchout Website\footnote{https://www.cosmos.esa.int/web/xmm-newton/sas-watchout-uvflux}. The Galactic extinction curve calculated by \citet{pei92} is used to convert $E(B-V)$ to extinction at another wavelength. 

\section{\xmm\ Data Analysis} \label{analysis}

In this section, we first introduce the EPIC-pn spectral analysis using a variable density disc reflection model. Second, we estimate the BH mass accretion rates by measuring the source flux at an optical band with OM. 

\begin{figure*}
\centering
\includegraphics[width=16cm]{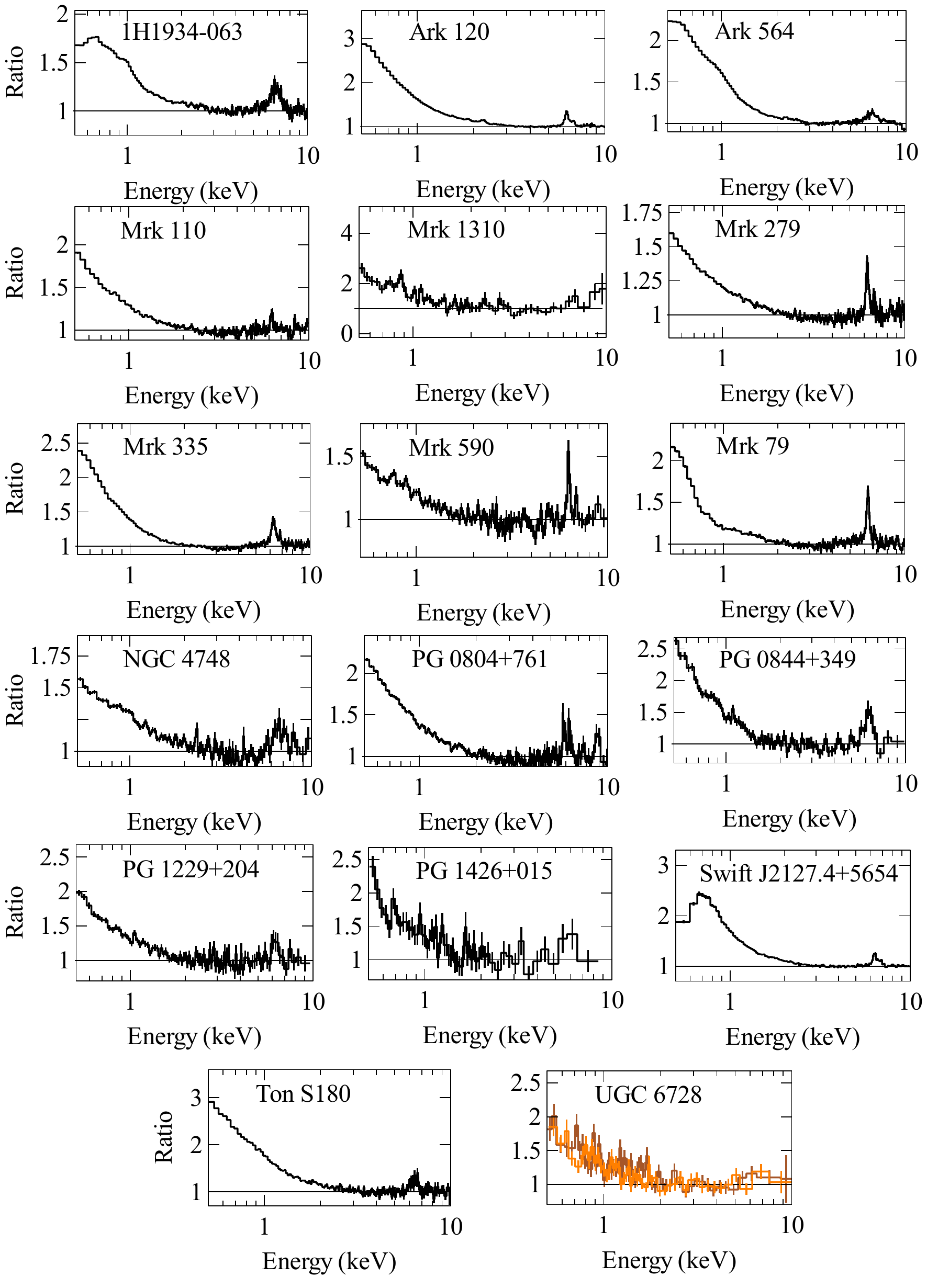}
\caption{Ratio plots for pn spectra fitted with absorbed power-law models. MOS1 (orange) and MOS2 (brown) spectra are shown for UGC~6728 as its pn observation is dominated by flaring particle background.}
\label{pic_fe}
\end{figure*}

\subsection{EPIC-pn Spectral Analysis and Disc Density Measurement} \label{pn_analysis}

The X-ray data analysis software XSPEC \citep{arnaud96} is used for broad band spectral analysis of EPIC-pn data. C-stat \citep{cash79} is used. The model \texttt{tbabs} in XSPEC is used to account for the Galactic absorption. Galactic column densities $N_{\rm H}$ calculated by \citet{willingale13} are used and fixed during the spectral fitting. The values of Galactic column density can be found in Table\,\ref{tab_bh_info}. 

First, we fit EPIC-pn spectra with an absorbed power-law model in the 2--3~keV and 8--10~keV band, ignoring the iron band and the soft excess. Corresponding ratio plots are shown in Figure\,\ref{pic_fe}. All the AGN in our sample show very strong soft excesses below 2 or 3~keV. Most AGN show evidence for emission features in the iron band. Some sources show a combination of broad and narrow emission lines. 

Second, we model the soft excess and the broad emission line in the iron band using the relativistic reflection model \texttt{relxilld} \citep[v1.2.0,][]{garcia16}. \texttt{relxilld} combines the convolution model \texttt{relconv} \citep{dauser13} and the illuminated ionised disc reflection model \texttt{xillverd} \citep{garcia16}. The \texttt{relconv} model calculates the relativistic effects and corresponding emissivity profiles for emission lines in the reflection spectrum. A broken power-law emissivity profile (inner index $q_{1}$, outer index $q_{2}$, threshold radius $R_{\rm r}$) is assumed. In cases where $q_{2}$ and $R_{\rm r}$ are not constrained, we assume a simple power-law emissivity profile instead. The disc reflection model \texttt{xillverd} allows the disc density parameter to vary between $\log(n_{\rm e})=$15--19. The solar abundance in \texttt{xillverd} is provided by \citet{grevesse96}. The ionisation parameter $\xi$ is defined as $\xi=4\pi F/n$ in unit of erg\,cm\,s$^{-1}$. The iron abundance ($Z_{\rm Fe}$) is a free parameter during our spectral fitting. The reflection model \texttt{xillver} in the same model package is used to account for the narrow emission line feature if shown in the iron band. A simple power-law model is used to model the coronal emission. The convolution model \texttt{cflux} is used to calculate the flux of each component in the 0.5--10~keV band. An empirical reflection fraction ($F_{\rm refl}/F_{\rm pl}$) is used for simplicity and future comparison with other reflection models \citep[e.g. \texttt{reflionx},][]{ross07}. In summary, the following models are used in XSPEC:
\begin{itemize}
    \item \texttt{tbabs * ( cflux * relxilld + cflux * powerlaw + cflux * xillver)}  (\texttt{MODEL1}) for sources that show narrow Fe K emission line in the iron band.
    \item \texttt{tbabs * ( cflux * relxilld + cflux * powerlaw )} (\texttt{MODEL2}) for sources that do not show any evidence for narrow Fe K emission line.
    \item \texttt{tbabs * ( cflux * relxilld + cflux * powerlaw + zgauss + zgauss)} (\texttt{MODEL3}) for sources that show complex ionised narrow emission lines in the iron band \citep[e.g. Ark~120,][]{matt14, nardini16}.
    \item \texttt{tbabs * ABSORBER * ( cflux * relxilld + cflux * powerlaw + cflux * xillver)} (\texttt{MODEL4}) for sources that show a thin warm absorber \citep[e.g. Mrk~335,][]{longinotti13,parker14} or a little obscuration \citep[e.g. Swift~J2127.4$+$5654,][]{miniutti09}. \texttt{ABSORBER} stands for the model that accounts for absorptions. For example, the ionised absorption model \texttt{warmabs} \citep{kallman01} is used to model warm absorber and \texttt{ztbabs} is used to model neutral obscuration.
\end{itemize}

\subsection{OM Photometry and Mass Accretion Rate} \label{om_analysis}

The same method as in \citet{raimundo12} is used to determine the BH mass accretion rate with \xmm\ OM observations, assuming a simple disc model with a steady accretion rate $\dot M$ and isotropic emission. The mass accretion rate is given by the following equation:
\begin{equation}
\dot M = 1.53\frac{\nu L_{\nu}}{10^{45}\cos(i)}^{3/2}\frac{10^{8}}{m_{\rm BH}} (M_{\odot}/{\rm yr})
\end{equation}
where $L_{\nu}$ is the luminosity in an optical band $\nu$ in erg\,s$^{-1}$ and $i$ is the viewing angle, which is obtained in the disc reflection analysis in Section\,\ref{pn_analysis}. As explained in \citet{raimundo12}, the lower energy band is less affected by a change of BH spin. For this reason, we choose B band (4500\AA) as the priority wavelength for the calculation, as in \citet{raimundo12}. If a source has no observation in B band in the \xmm\ archive, we choose the lowest energy band for calculation. Notice that $L_{\nu}$ has been corrected for Galactic extinction using the Galactic extinction curve calculated by \citet{pei92} for each source. Table \ref{tab_om} shows the observed flux $F_{\nu}$ in the optical band named in the second column and corresponding mass accretion rate $\dot{m}=\dot{M}/\dot M_{\rm Edd}$, where $\dot M_{\rm Edd}$ is the Eddington accretion rate. 

Note that we did not calculate the accretion rate by calculating the bolometric luminosity due to the large uncertainty of the bolometric conversion factor for the X-ray band flux \citep{vasudevan07} and the accretion efficiency in AGN \citep{raimundo12}.  

\begin{table}
\caption{The flux of each source in an optical band measured by performing photometry with \xmm\ OM observations. $F_{\nu}$ is the observed flux in the unit of $10^{-16}$\,erg\,s$^{-1}$\,cm$^{-2}$\,\AA$^{-1}$ in the band shown in the second column. The mass accretion rate $\dot m$ is in units of the Eddington accretion rate.}
\label{tab_om}
\centering
\begin{tabular}{ccccc}
\hline\hline
Name & Band & $F_{\nu}$  & $\dot{m}$ \\
\hline
1H1934 & U & $59.8\pm0.3$ & $1.1^{+2.3}_{-0.6}$\\
Ark~120 & V & $110.9\pm0.7$ & $0.76^{+0.12}_{-0.08}$\\
Ark~564 & UVW1 & $71.7\pm0.6$ & $1.7^{+1.7}_{-1.1}$\\
Mrk~110 & B & $72.8\pm0.4$  & $0.90^{+0.23}_{-0.19}$ \\
Mrk~1310 & U & $14.88\pm0.13$  & $0.6^{+0.14}_{-0.10}$\\
Mrk~279 & U & $144.6\pm0.7$ & $0.75^{+0.27}_{-0.16}$\\
Mrk~335 & B & $88.3\pm0.5$ & $0.74^{+0.08}_{-0.07}$\\
Mrk~590 & B & $50.7\pm0.4$ & $0.31^{+0.06}_{-0.05}$\\
Mrk~79 & B & $52.0\pm0.3$  & $0.13^{+0.05}_{-0.07}$\\
NGC~4748 & U & $81.5\pm0.3$  & $2.2^{+1.2}_{-0.5}$ \\
PG~0804 & V & $105.3\pm0.3$  & $1.13^{+0.15}_{-0.12}$ \\
PG~0844 & B & $65.7\pm0.4$  & $1.20^{+1.0}_{-0.4}$\\
PG~1229 & UVM2 & $69.9\pm0.8$  & $0.5^{+0.4}_{-0.2}$\\
PG~1426 & UVM2 & $302.9\pm1.7$ & $0.28^{+0.13}_{-0.06}$ \\
Swift~J2127 & B & $2.43\pm0.06$ & $1.0^{+0.5}_{-0.2}$\\
Ton~S180 & U & $49.3\pm0.5$ & $6^{+12}_{-5}$\\
UGC~6728 & U & $53.7\pm0.4$ & $0.58^{+0.76}_{-0.21}$\\
\hline\hline
\end{tabular}
\end{table}

\begin{figure*}
\centering
\includegraphics[width=16cm]{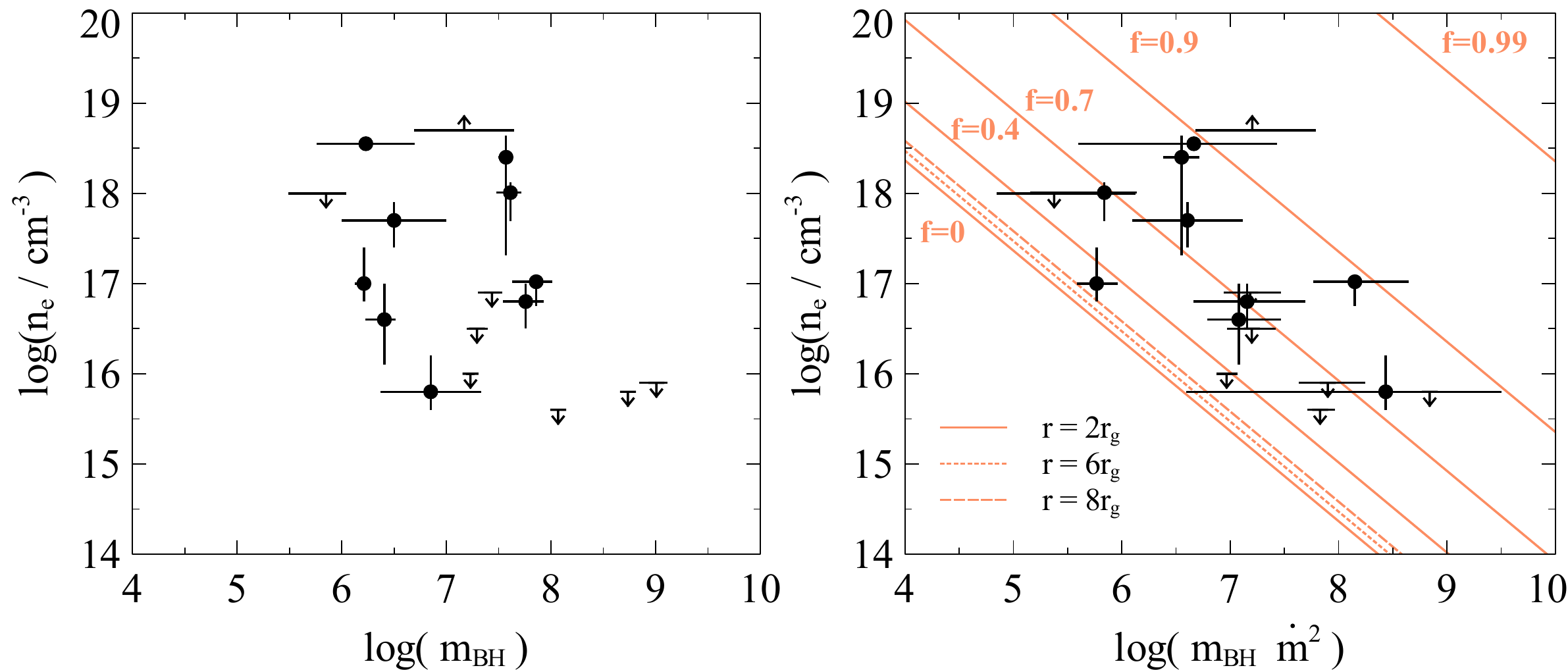}
\caption{Left: Disc density $\log(n_{\rm e})$ versus BH mass $\log(m_{\rm BH})$. Only upper limits of disc density are obtained for black holes with $\log(m_{\rm BH})>8$, indicating a lower disc density in high-$m_{\rm BH}$ AGN. Right: Disc density $\log(n_{\rm e})$ versus $\log(m_{\rm BH} \dot m ^{2})$. The solid orange lines are the solutions for disc density at $r=2$\,$r_{\rm g}$ for different values of $f$ \citep{svensson94}. The inner radius is assumed to be at $R_{\rm ISCO}$ of a maximally spinning BH ($R_{\rm in}=1r_{\rm g}$). The dotted and dashed straight lines are for $r=$6, 8\,$r_{\rm g}$ and $f=0$.}
\label{pic_ne1}
\end{figure*}

\section{Results} \label{results}

Appendix\,\ref{re1} presents the details of EPIC-pn spectral analysis for individual sources in our sample. The best-fit parameters are reported in Table A2 and the best-fit models with corresponding best-fit ratio plots are shown in Fig.\,\ref{pic_final1}, \ref{pic_final2}, and \ref{pic_final3}. In this section, we summarise the results of the disc density measurements and compare our results with previous work.

\subsection{Disc Denstities in Seyfert 1 Galaxies}

We present the best-fit disc density values versus BH masses in the left panel of Fig.\,\ref{pic_ne1}. No obvious correlation between disc densities and BH masses in our sample is found. However, only upper limits of disc densities are achieved for sources with $\log(m_{\rm BH})>8$. This matches the previous analysis of another Seyfert 1 galaxy 1H~0419$-$577 where the mass of the central BH is estimated to be $\log(m_{\rm BH})\approx8$ and a disc density of $\log(n_{\rm e})<15.2$ is obtained with 90\% confidence \citep{jiang18d}. Disc densities significantly higher than $\log(n_{\rm e})=15$ are found in AGN with $\log(m_{\rm BH})<7$. By comparing the best-fit disc densities with BH masses, we find that the assumption for a fixed disc density of $\log(n_{\rm e})=15$ is mostly appropriate for SMBH with $\log(m_{\rm BH})>8$ while a larger disc density is required for SMBH with $\log(m_{\rm BH})<7$.

A second factor that changes the disc density is the BH mass accretion rate. \citet[][]{shakura73} predicts that the disc density of a radiation pressure-dominated disc and the mass accretion rate has the following relation \red{$\log(n_{\rm e})\propto-2\log(\dot m)$.} The right panel of Fig.\,\ref{pic_ne1} presents the disc density solution given by Eq.8 in \citet{svensson94}, where $f$ is the fraction of power released from the disc to the corona. When $f=0$, this solution reproduces the results in \citet{shakura73}. The solid lines show the disc densities at $R=2R_{\rm g}$ for different $f$, assuming: 1) a maximum BH spin; 2) the inner radius of the disc is $R_{\rm ISCO}$; 3) the conversion factor from radiative pressure to radiation flux in the radiative diffusion equation $\xi^{\prime}=1$ \footnote{The prime symbol is to distinguish the conversion factor from the disc ionisation parameter $\xi$.}. The dotted and dashed lines show the disc densities at $R=6$, $8R_{\rm g}$ for $f=0.01$. The radius $R$ has less impact on the disc density than $f$, as shown in Fig.\,\ref{pic_ne1}. 

We present our results in the $\log(n_{\rm e})$ vs. $\log(m_{\rm BH} \dot m ^{2})$ diagram in the right panel of Fig.\,\ref{pic_ne1}. There is tentative evidence for an anti-correlation between $\log(n_{\rm e})$ and $\log(m_{\rm BH} \dot m ^{2})$. We use a Monte-Carlo approach to estimate the significance of the correlation. We assume $\log(n_{\rm e})$ and $\log(m_{\rm BH} \dot m ^{2})$ follow normal distributions. 100,000 sets of points are drawn from distributions with the same mean (mean $\log(m_{\rm BH} \dot m ^{2})=7.1$, $\log(n_{\rm e})=16.89$) and deviation ($\sigma_{\log(m_{\rm BH} \dot m ^{2})}=0.92$, $\sigma_{\log(n_{\rm e})}=1.17$) of the sample. We find 11304 sets of points that exceed the Spearman correlation coefficient of our result (-0.67). This gives 11\% probability of the observed correlation from randomly distributed points. By considering the large uncertainties of mass measurements, the significance of the correlation shown in Fig.\,\ref{pic_ne1} is even lower. The correlation between these two parameters is weak in our sample due to the following reasons: 1) the uncertainties of the BH mass measurements; 2) other physics, such as the vertical structure of the disc density and the fraction ($f$) of power that is released from the disc to the coronal region. We will discuss the effect of $f$ on the disc density in Section\,\ref{discuss}. 

In summary, we find some evidence for a trend that high disc density is commonly seen in AGN that show low $m_{\rm BH} \dot m ^{2}$ values. It is important to note that the Galactic BH GX~339-4 shows a similar pattern. The disc density of GX~339$-$4 in the low flux-hard state is higher the density in the high flux-soft state \citep{jiang19}.

\subsection{Comparison with Previous Work}

\subsubsection{Disc Iron Abundances}

\begin{figure}
\centering
\includegraphics[width=\columnwidth]{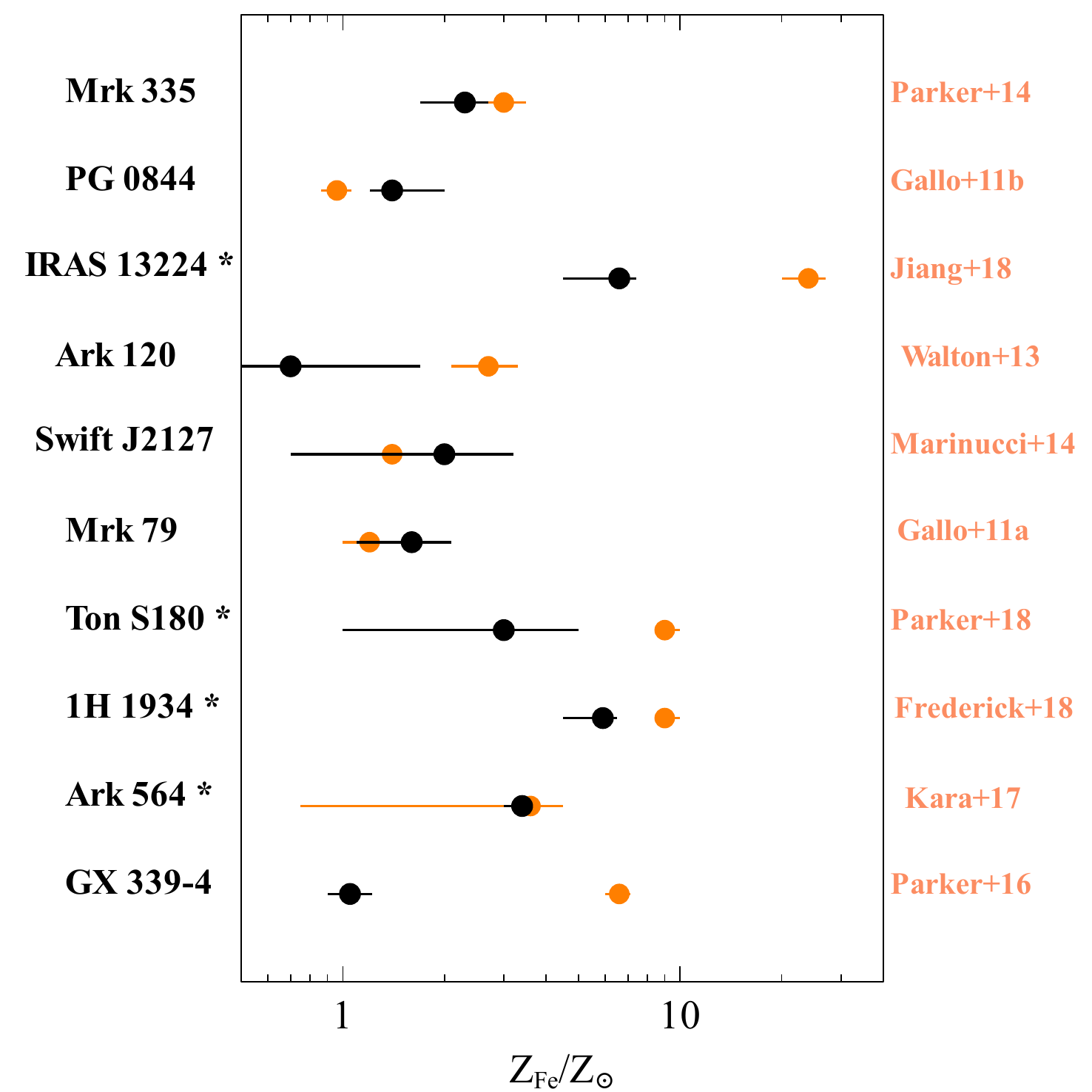}
\caption{Best-fit iron abundances $Z_{\rm Fe}$ obtained by modelling with a variable density model shown in black compared with previous results shown in orange where $\log(n_{\rm e})=15$ was assumed. Corresponding references are labelled on the right side. * An additional model component is used for soft excess in previous analyses for these AGN.}
\label{pic_ze}
\end{figure}

A super solar iron abundance was often seen in previous reflection-based spectral analysis where a fixed disc density ($n_{\rm e}=10^{15}$\,cm$^{-3}$) was assumed. A significant decrease of inferred iron abundance is found in our new disc reflection modelling. Fig.\,\ref{pic_ze} presents the best-fit disc iron abundances in this work shown in black and the best-fit iron abundances in previous work in orange. Corresponding references are labelled on the right side. This figure only includes the sources that have reflection analyses in previous published work. The disc iron abundances obtained by using a high density model for IRAS~13224 and GX~339$-$4 are from \citet{jiang18,jiang19}.   

Several sources in our sample show a significant decrease of inferred disc iron abundance when a high density disc reflection model is used. These sources are 1H~1934, Ton~S180, IRAS~13224 and GX~339$-$4. A similar conclusion was found in \citet{tomsick18}, where an iron abundance at the solar level is required for the intermediate state of the BH XRB Cyg~X-1 when a high density model is used. Only PG~0844 shows a slightly higher disc iron abundance compared to previous results. However the abundance is still near the solar level ($Z_{\rm Fe}<3Z_{\odot}$).

\begin{figure*}
    \centering
    \includegraphics[width=16cm]{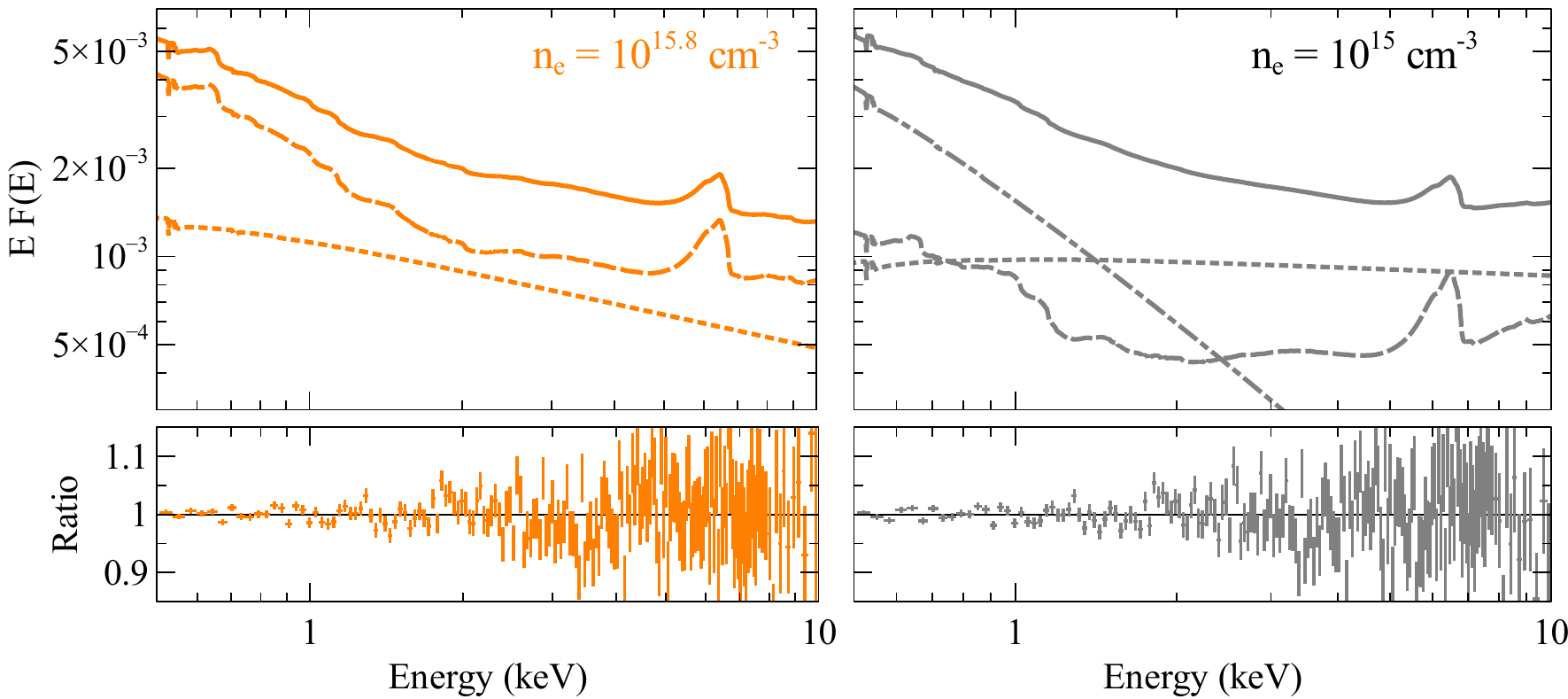}
    \caption{Left: The best-fit models for Ton~S180 when a high density disc model is used and the corresponding ratio plot. Solid: total model; dashed: high density disc model with $\log(n_{\rm e})=15.6$; dotted: power-law continuum. The unit of the y-axis of the upper panel is keV\,cts\,cm$^{-2}$\,s$^{-1}$. Right: Same as the left panel but an additional soft \texttt{cutoff} model (dash-dotted line) is used to model the soft excess. A fixed disc density of $\log(n_{\rm e})=15$ is assumed. In this model, the coronal emission shows a harder continuum and the reflection component makes less contribution to the total emission compared to the models in the left panel.}
    \label{pic_180}
\end{figure*}

The decrease of the inferred iron abundances is due to the increase of the continuum in the reflection component when a high density model is used. For example, the best-fit reflection fraction for Ton~S180 is $f_{\rm refl}=1.9\pm0.7$ when the spectrum is modelled using a variable density disc reflection model \footnote{Note that the reflection fraction $f_{\rm refl}$ here is not the same reflection fraction used in \citet{dauser16}. An empirical definition of $F_{\rm refl}/F_{\rm pl}$ in the \xmm\ band is used in order to quantify the relative strength of the reflection component compared to the coronal emission.}. The averaged spectrum of Ton~S180 is dominated by the disc reflection component and requires an iron abundance of $Z_{\rm Fe}=3\pm2 Z_{\odot}$. Instead of using a high density model, we fit the soft excess emission of Ton~S180 with a soft \texttt{cutoff} model in addition to a $\log(n_{\rm e})=15$ disc model, similar to the work of \citet{parker18}. Although such a model combination slightly improves the fit by $\Delta$C-stat=7 with 1 more parameter, the disc reflection component with a fixed disc density requires a very high iron abundance of $Z_{\rm Fe}>8Z_{\odot}$. See Fig.\,\ref{pic_180} for comparison of best-fit models for disc densities and Table\,\ref{tab_tons180_cutoff} for best-fit parameters when \texttt{cutoff} is used to model the soft excess. A much lower reflection fraction of $f_{\rm refl}=0.70$ is required. Therefore in order to model the strong Fe~K emission feature in the iron band of Ton~S180, a higher iron abundance is required.

Although a high density disc model is able to decrease the inferred iron abundance, some AGN still show a high iron abundance compared to solar \citep[e.g. IRAS~13224$-$3809,][]{jiang18}. In future, we will further study the metallicity of these systems by allowing more element abundances to be free parameters during the spectral modelling.  

\subsubsection{BH Spins and Disc Inclination Angles}

\begin{figure*}
\centering
\includegraphics[width=\columnwidth]{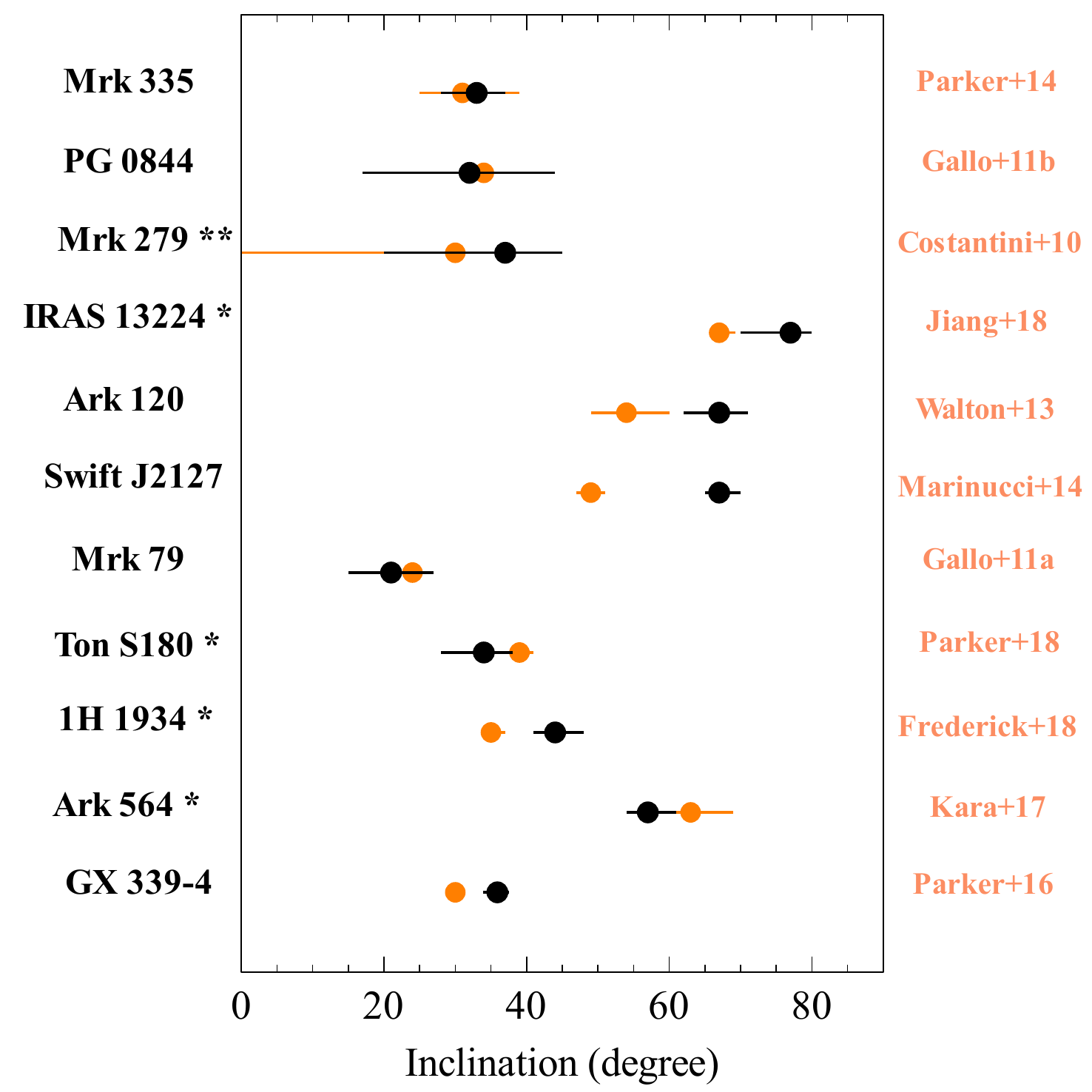}
\includegraphics[width=\columnwidth]{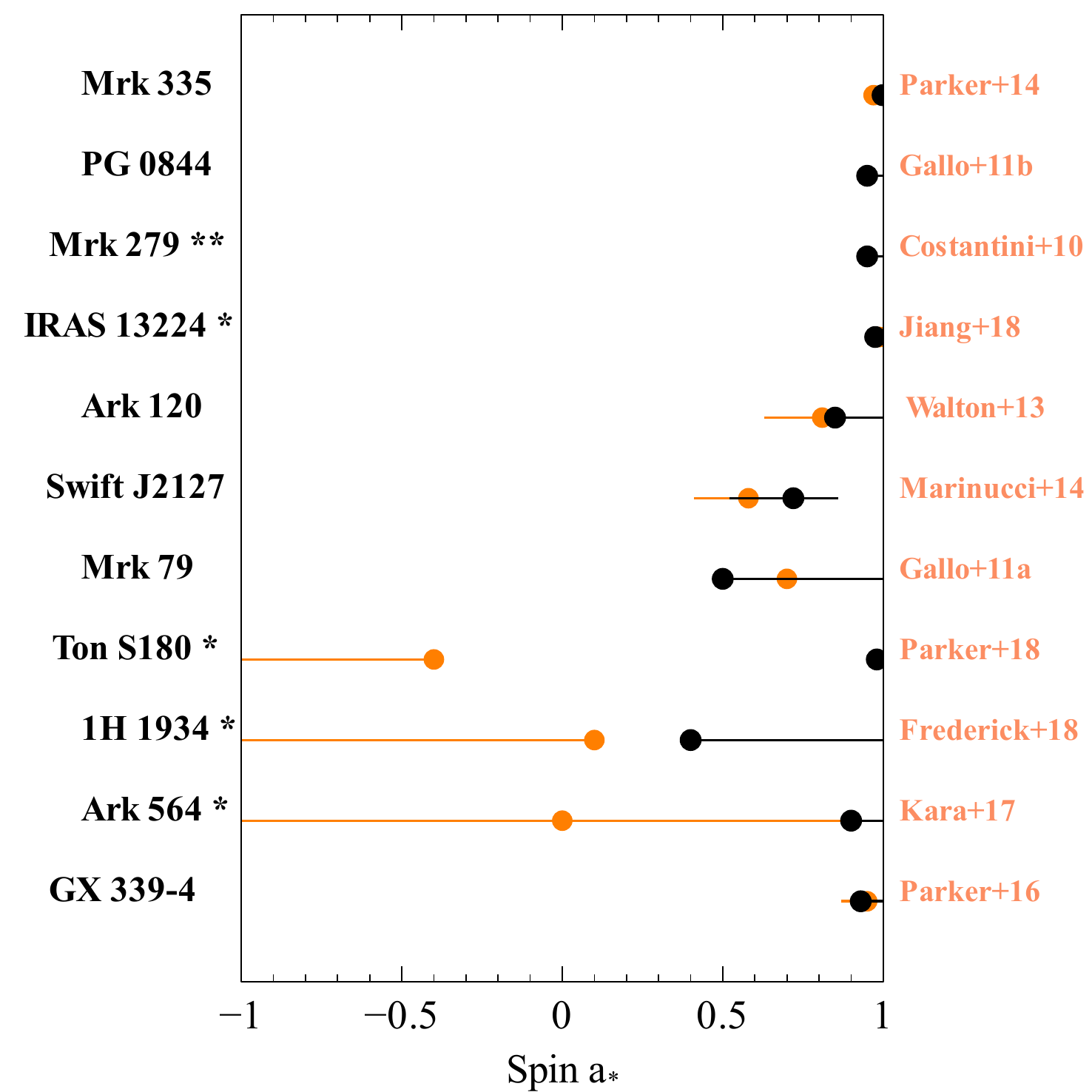}
\caption{Left: Comparison of disc inclination angles obtained by modelling with a variable density model shown in black and previous results shown in orange where $\log(n_{\rm e})=15$ was assumed. Right: Comparison of BH spins. * An additional model component is used for soft excess in previous analyses for these AGN. ** \citet{costantini10} only obtained the inclination angle of Mrk~279 by modelling the broad iron line with the disc emission line model \texttt{laor}, assuming a maximum BH spin.}
\label{pic_spin_i}
\end{figure*}

Fig.\ref{pic_spin_i} shows disc inclination angles and BH spins obtained in our work in black and previous analyses in orange. Although we do not expect the orientation of an AGN and the spin of a SMBH to vary in different observations, we can not rule out  instrumental effects of different observations on the measurements \citep[e.g. see discussion section of][]{brenneman13}. Moreover, the X-ray continuum emission from the Seyfert AGN mentioned in Fig.\,\ref{pic_spin_i} is known to be very variable \citep[e.g.][]{frederick18}. The continuum modelling can potentially have an impact on the determination of the red and blue wings of their broad Fe~K$\alpha$ emission lines and thus affect the measurements of BH spin and disc inclination angle respectively. 

Most of the disc inclination angles measured using high density disc reflection spectroscopy are consistent with previous results over the 90\% confidence ranges, except for 1H~1934, Ark~120, Swift~J2127, IRAS~13224, and GX~339$-$4. We only obtain an upper limit of $n_{\rm e}$ for Ark~120. Therefore the difference of inclination angles for Ark~120 is unlikely due to the high density model. The other four sources all show higher inferred inclination angles when a variable density model is used. Although the disc inclination angle of GX~339$-$4 using a high disc density model is higher than the result in \citet{jingyi18} and lower than the result in \citet{parker16}, our measurement is still consistent with previous reflection-based analyses within a 2$\sigma$ uncertainty range. See the discussion section of \citet{jiang19}. We also notice that \citet{tomsick18} found a different conclusion when using a high density model for Cyg~X$-$1. A significantly smaller disc inclination angle is found for this source when a variable density model is used. 

Most of the best-fit BH spins for our sample given by high density disc reflection spectroscopy are near the Kerr limit and consistent with previous analyses within 90\% confidence ranges, except for 1H~1934 and Ton~S180. In previous analyses of 1H~1934 and Ton~S180 \citep{frederick18,parker18}, an additional component was used to model the soft excess. In the previous section, we explain how a high density model changes the continuum modelling. For example, the power-law continuum of Ton~S180 has $\Gamma=2.38$ when a high density model is used. By contrast, a harder continuum of $\Gamma=2$ is found for the same spectrum if a \texttt{cutoff} model is added for soft excess. See Fig.\,\ref{pic_180} for comparison of two models. In this case, the spin is not constrained (see Table\,\ref{tab_tons180_cutoff}). By contrast, \citet{nardini12} obtained a disc inner radius of $R_{\rm in}\approx2.4r_{\rm g}$ by modelling the soft excess of Ton~S180 with a second relativistic disc reflection model instead of a \texttt{cutoff} model and assuming $\log(n_{\rm e})=15$. Their disc inner radius measurement is consistent with our spin measurement using a high density disc reflection model.

Similarly, \citet{kara17} claims the spin of Ark~564 is not constrained by analysing its \suzaku\ and \nustar\ spectra while \citet{walton13} obtains a spin of $a_{*}=0.96^{+0.01}_{-0.06}$ for the same source by analysing its \suzaku\ observations. The latter measurement is consistent with our result. The difference in \citet{kara17} was that an additional thermal bremsstrahlung component was used to model the soft excess and the \suzaku\ XIS spectrum below 1~keV was ignored. This work and \citet{walton13} fit the soft excess emission with disc reflection model.

Future simultaneous hard X-ray observations, such as from \nustar, will be helpful for constraining the continuum emission by including the reflection Compton hump above 10~keV. An example is the analysis of the simultaneous \nustar\ and \xmm\ observation of the NLS1 Mrk~1044 in \citet{mallick18}, where a high density disc reflection model of $\log(n_{\rm e})\approx16$ explains the 0.5--78~keV band spectrum of this source with no requirement for an additional component.

\section{Discussion and Conclusions} \label{discuss}

\begin{figure}
\centering
\includegraphics[width=\columnwidth]{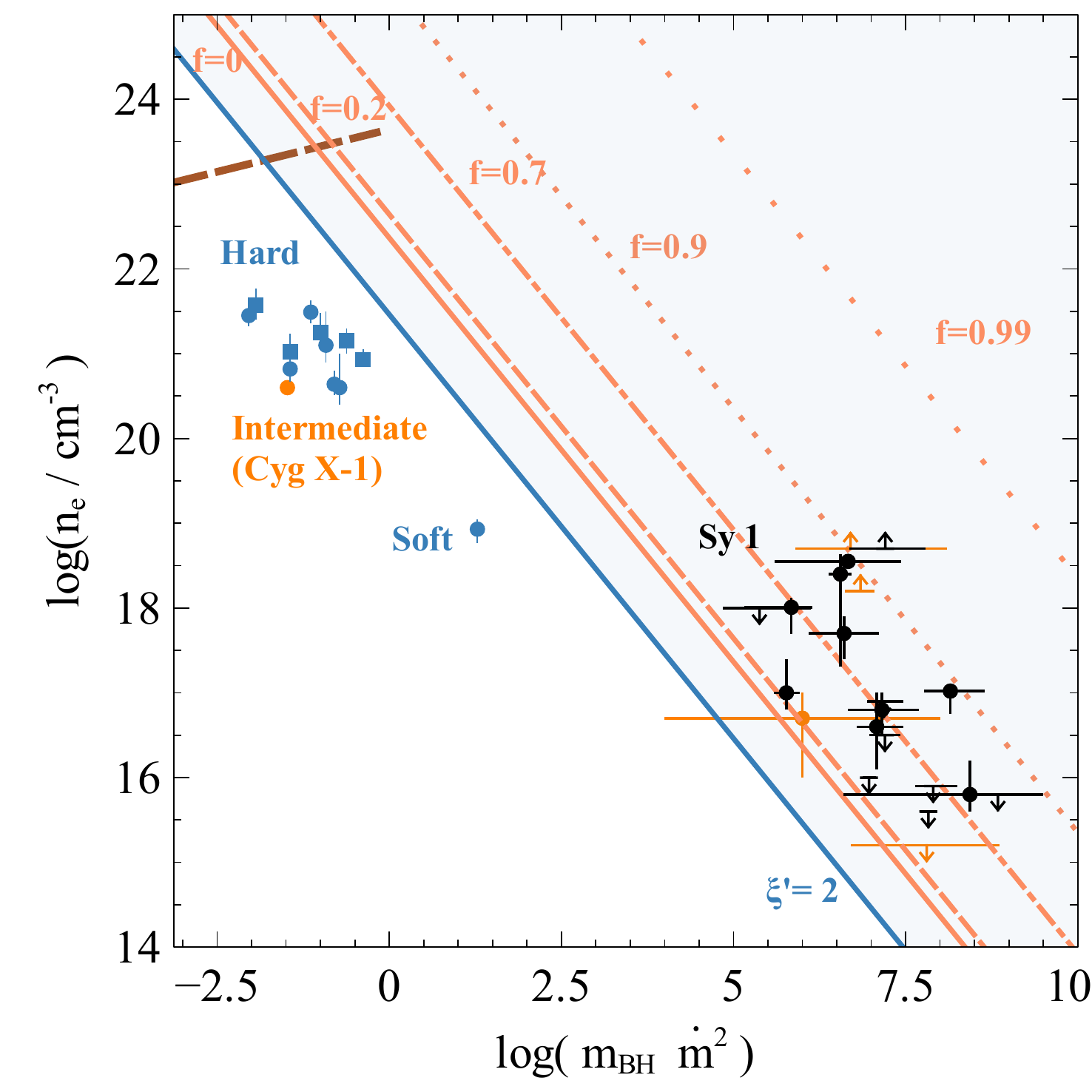}
\caption{Disc density $\log(n_{\rm e})$ versus $\log(m_{\rm BH} \dot m ^{2})$ for our sample and previous analysis. Previous analysis are marked by orange points (see text for references). Blue circles represent GX~339$-$4 observations in 2015 and squares represent observations in 2013 that are analysed in \citet{jiang19}. A BH mass of $m_{\rm BH}=10$ is assumed for GX339$-$4. Orange lines are the same as the ones in Fig.\,\ref{pic_ne1}. Blue solid line is the solution assuming $f=0$, $\xi^{\prime}=2$, and $r=2r_{\rm g}$. The red dashed line shows the solution for a gas pressure-dominated disc at $r=2r_{\rm g}$ assuming $f=0$. All the disc density curves are calculated by \citet{svensson94}.}
\label{pic_ne2}
\end{figure}

We summarise the disc density measurements in this paper and previous analyses \citep{tomsick18, mallick18, garcia18, jiang18, jiang19} in Fig.\,\ref{pic_ne2}. The results from previous work of AGN are marked by orange points. Blue circles are for GX~339-4 observations in 2015 and blue squares are for observations in 2013. Black points are the measurements for our sample. The orange straight lines are the same as the ones as in Fig.\,\ref{pic_ne1} and the blue straight line is the solution for $f=0$, $R=2R_{\rm g}$, and $\xi^{\prime}=2$. We can draw the following conclusions from this diagram:

First, a significantly higher disc density is found in stellar-mass BHs than in SMBHs. \red{65\% of SMBHs in our sample show evidence for a disc density significantly higher than $\log(n_{\rm e})=15$.} In the hard state of GX~339-4 and the intermediate state of Cyg~X-1, the density of the disc is at least $\log(n_{\rm e})>20.5$. In the high flux state of GX~339-4, a disc density of $\log(n_{\rm e})\approx19$ is required, similar with the disc density in AGN with $\log(m_{\rm BH} \dot m ^{2})<7$. 

Second, the accretion rate affects the disc density in the same way as the BH mass: a higher disc density is found where the BH accretes at a lower fraction of its Eddington limit. This conclusion has been found previously by studying the different states of the BH XRB GX~339$-$4 \citep{jiang19} where only the accretion rate can be changing. Similarly, we find tentative evidence for a similar conclusion for AGN. For example, Mrk~509 and PG~0804$+$761 have a similar BH mass of $m_{\rm BH}\approx10^{8}$ \citep{bentz15,garcia18}. However, Mrk~509 has a higher disc density than PG~0804$+$761. \red{This might be due to the higher accretion rate in PG~0804$+$761 ($\dot{m}\approx1.0-1.3$, see Table \ref{tab_om})} than in Mrk~509 \citep[$\dot{m}=0.2-0.4$,][]{petrucci13}.

Third, theoretically most coronal heating mechanisms assume that a large fraction of the disc energy is dissipated in the coronal region \citep[e.g. magnetic coronae model,][]{  galeev79, coroniti81, stella84}. Assuming $\xi^{\prime}=1$, our analysis suggests at least 10\% of the power in the disc is released to the corona in AGN. If a higher value of $\xi^{\prime}$ \citep[e.g. $\xi^{\prime}=2$,][]{svensson94} is assumed, an even higher $f$ is expected. See the blue and orange solid lines in Fig.\ref{pic_ne2}.

Fourth, although there is a weak correlation between $\log(n_{\rm e})$ and $\log(m_{\rm BH} \dot m ^{2})$ in our sample, a disc density of $\log(n_{\rm e})>16$ is clearly found in AGN with $\log(m_{\rm BH} \dot m ^{2})<6.5$. The weak correlation could be due to other uncertain effects, such as different $f$, the vertical structure of the disc density in reality, or the uncertainties of the BH mass measurements.

Fifth, the disc densities that our reflection model obtains for BH XRBs are significantly lower than the prediction of the standard thin disc model \citep[the solid orange line in Fig.\,\ref{pic_ne2},][]{shakura73}. Some potential explanations are: 1) the disc density parameter in the reflection model is the density in the surface of the disc while the thin disc models assume a uniform disc density in the vertical direction; 2) the BH mass and the distance measurements of GX~339$-$4 are uncertain. However, we notice that the disc density of Cyg~X-1 is still below the prediction, although the mass and the distance of Cyg~X$-$1 are well constrained in this case \citep{orosz11,reid11}; 3) the stellar-mass BH discs during the observations considered in \citet{tomsick18,jiang19} are likely to be dominated by gas pressure instead of radiation pressure with $L_{\rm Bol} \approx L_{\rm X}<4\%L_{\rm Edd}$. At such low luminosities, they correspond to the gas pressure-dominated disc regime according to \citet{svensson94}. Consequences of gas pressure-dominated discs happening during the state transition of BH transients will be addressed in future work. However, the densities measured in XRBs are still 10--100 times lower than the density of a gas pressure-dominated disc in \citet{shakura73}, which is shown by the red dashed line in Fig.\,\ref{pic_ne2}. It suggests that more physics is needed to be considered, such as the vertical structure of a gas pressure-dominated disc. In Appendix\,\ref{gas}, we estimate the disc density within optical depth $\tau<1$ of a gas pressure-dominated disc.

In conclusion, we find that the high density disc reflection model can not only decrease the inferred iron abundance but also successfully explains the 0.5--10~keV band spectra of the Seyfert 1 galaxies in our sample with no requirement for additional components for the soft excess emission. The density of the disc is significantly higher than the previous $\log(n_{\rm e})=15$ assumption in AGN with $\log(m_{\rm BH}\dot{m}^{2})<7.5$ in our sample. This is consistent with the prediction of the standard thin disc model \citep{shakura73,svensson94}. More future work is needed to complete the high density disc reflection spectroscopy for AGN in following approaches: 1) abundances of other elements in addition to iron need to be considered; 2) simultaneous broad band observations of these Seyfert galaxies are required (e.g. from \textit{NuSTAR}) to test whether the high density model can account for the Compton hump above 10~keV; 3) time-resolved spectral analysis is required to study the short-term variability of the disc density within one observation.

\section*{Acknowledgements}

J.J. acknowledges support by the Cambridge Trust and the Chinese Scholarship Council Joint Scholarship Programme (201604100032). D.J.W. acknowledges support from an STFC Ernest Rutherford Fellowship. A.C.F. acknowledges support by the ERC Advanced Grant 340442. M.L.P. is supported by European Space Agency (ESA) Research Fellowships. J.A.G. acknowledges support from the Alexander von Humboldt Foundation. The authors are also grateful to Andrew J. Young, Paul C. Hewett and Sergei Dyda for valuable discussion.

%%%%%%%%%%%%%%%%%%%%%%%%%%%%%%%%%%%%%%%%%%%%%%%%%%



%\bibliographystyle{mnras}

\bibliographystyle{mnras}
\bibliography{highne.bib} % if your bibtex file is called example.bib

\begin{thebibliography}{}
\makeatletter
\relax
\def\mn@urlcharsother{\let\do\@makeother \do\$\do\&\do\#\do\^\do\_\do\%\do\~}
\def\mn@doi{\begingroup\mn@urlcharsother \@ifnextchar [ {\mn@doi@}
  {\mn@doi@[]}}
\def\mn@doi@[#1]#2{\def\@tempa{#1}\ifx\@tempa\@empty \href
  {http://dx.doi.org/#2} {doi:#2}\else \href {http://dx.doi.org/#2} {#1}\fi
  \endgroup}
\def\mn@eprint#1#2{\mn@eprint@#1:#2::\@nil}
\def\mn@eprint@arXiv#1{\href {http://arxiv.org/abs/#1} {{\tt arXiv:#1}}}
\def\mn@eprint@dblp#1{\href {http://dblp.uni-trier.de/rec/bibtex/#1.xml}
  {dblp:#1}}
\def\mn@eprint@#1:#2:#3:#4\@nil{\def\@tempa {#1}\def\@tempb {#2}\def\@tempc
  {#3}\ifx \@tempc \@empty \let \@tempc \@tempb \let \@tempb \@tempa \fi \ifx
  \@tempb \@empty \def\@tempb {arXiv}\fi \@ifundefined
  {mn@eprint@\@tempb}{\@tempb:\@tempc}{\expandafter \expandafter \csname
  mn@eprint@\@tempb\endcsname \expandafter{\@tempc}}}

\bibitem[\protect\citeauthoryear{{Arnaud}}{{Arnaud}}{1996}]{arnaud96}
{Arnaud} K.~A.,  1996, {XSPEC: The First Ten Years}

\bibitem[\protect\citeauthoryear{{Arnaud} et~al.,}{{Arnaud}
  et~al.}{1985}]{arnaud85}
{Arnaud} K.~A.,  et~al., 1985, \mn@doi [\mnras] {10.1093/mnras/217.1.105},
  \href {http://adsabs.harvard.edu/abs/1985MNRAS.217..105A} {217, 105}

\bibitem[\protect\citeauthoryear{{Bentz} \& {Katz}}{{Bentz} \&
  {Katz}}{2015}]{bentz15}
{Bentz} M.~C.,  {Katz} S.,  2015, \mn@doi [\pasp] {10.1086/679601}, \href
  {http://adsabs.harvard.edu/abs/2015PASP..127...67B} {127, 67}

\bibitem[\protect\citeauthoryear{{Bianchi}, {Guainazzi}, {Matt}, {Fonseca
  Bonilla}  \& {Ponti}}{{Bianchi} et~al.}{2009}]{bianchi09}
{Bianchi} S.,  {Guainazzi} M.,  {Matt} G.,  {Fonseca Bonilla} N.,   {Ponti} G.,
   2009, \mn@doi [\aap] {10.1051/0004-6361:200810620}, \href
  {http://adsabs.harvard.edu/abs/2009A%26A...495..421B} {495, 421}

\bibitem[\protect\citeauthoryear{{Boller}, {Balestra}  \&
  {Kollatschny}}{{Boller} et~al.}{2007}]{boller07}
{Boller} T.,  {Balestra} I.,   {Kollatschny} W.,  2007, \mn@doi [\aap]
  {10.1051/0004-6361:20066343}, \href
  {http://adsabs.harvard.edu/abs/2007A&A...465...87B} {465, 87}

\bibitem[\protect\citeauthoryear{{Brenneman}}{{Brenneman}}{2013}]{brenneman13}
{Brenneman} L.,  2013, {Measuring the Angular Momentum of Supermassive Black
  Holes}, \mn@doi{10.1007/978-1-4614-7771-6.
}

\bibitem[\protect\citeauthoryear{{Cash}}{{Cash}}{1979}]{cash79}
{Cash} W.,  1979, \mn@doi [\apj] {10.1086/156922}, \href
  {https://ui.adsabs.harvard.edu/\#abs/1979ApJ...228..939C} {228, 939}

\bibitem[\protect\citeauthoryear{{Coroniti}}{{Coroniti}}{1981}]{coroniti81}
{Coroniti} F.~V.,  1981, \mn@doi [\apj] {10.1086/158739}, \href
  {http://adsabs.harvard.edu/abs/1981ApJ...244..587C} {244, 587}

\bibitem[\protect\citeauthoryear{{Costantini}, {Kaastra}, {Korista}, {Ebrero},
  {Arav}, {Kriss}  \& {Steenbrugge}}{{Costantini} et~al.}{2010}]{costantini10}
{Costantini} E.,  {Kaastra} J.~S.,  {Korista} K.,  {Ebrero} J.,  {Arav} N.,
  {Kriss} G.,   {Steenbrugge} K.~C.,  2010, \mn@doi [\aap]
  {10.1051/0004-6361/200912555}, \href
  {https://ui.adsabs.harvard.edu/\#abs/2010A&A...512A..25C} {512, A25}

\bibitem[\protect\citeauthoryear{{Dauser}, {Garcia}, {Wilms}, {B{\"o}ck},
  {Brenneman}, {Falanga}, {Fukumura}  \& {Reynolds}}{{Dauser}
  et~al.}{2013}]{dauser13}
{Dauser} T.,  {Garcia} J.,  {Wilms} J.,  {B{\"o}ck} M.,  {Brenneman} L.~W.,
  {Falanga} M.,  {Fukumura} K.,   {Reynolds} C.~S.,  2013, \mn@doi [\mnras]
  {10.1093/mnras/sts710}, \href
  {http://adsabs.harvard.edu/abs/2013MNRAS.430.1694D} {430, 1694}

\bibitem[\protect\citeauthoryear{{Dauser}, {Garc{\'{\i}}a}, {Walton},
  {Eikmann}, {Kallman}, {McClintock}  \& {Wilms}}{{Dauser}
  et~al.}{2016}]{dauser16}
{Dauser} T.,  {Garc{\'{\i}}a} J.,  {Walton} D.~J.,  {Eikmann} W.,  {Kallman}
  T.,  {McClintock} J.,   {Wilms} J.,  2016, \mn@doi [\aap]
  {10.1051/0004-6361/201628135}, \href
  {http://adsabs.harvard.edu/abs/2016A%26A...590A..76D} {590, A76}

\bibitem[\protect\citeauthoryear{{De Marco}, {Ponti}, {Cappi}, {Dadina},
  {Uttley}, {Cackett}, {Fabian}  \& {Miniutti}}{{De Marco}
  et~al.}{2013}]{demarco13}
{De Marco} B.,  {Ponti} G.,  {Cappi} M.,  {Dadina} M.,  {Uttley} P.,  {Cackett}
  E.~M.,  {Fabian} A.~C.,   {Miniutti} G.,  2013, \mn@doi [\mnras]
  {10.1093/mnras/stt339}, \href
  {https://ui.adsabs.harvard.edu/\#abs/2013MNRAS.431.2441D} {431, 2441}

\bibitem[\protect\citeauthoryear{{Fabian} et~al.,}{{Fabian}
  et~al.}{2009}]{fabian09}
{Fabian} A.~C.,  et~al., 2009, \mn@doi [\nat] {10.1038/nature08007}, \href
  {http://adsabs.harvard.edu/abs/2009Natur.459..540F} {459, 540}

\bibitem[\protect\citeauthoryear{{Frank}, {King}  \& {Raine}}{{Frank}
  et~al.}{2002}]{frank02}
{Frank} J.,  {King} A.,   {Raine} D.~J.,  2002, {Accretion Power in
  Astrophysics: Third Edition}

\bibitem[\protect\citeauthoryear{{Frederick}, {Kara}, {Reynolds}, {Pinto}  \&
  {Fabian}}{{Frederick} et~al.}{2018}]{frederick18}
{Frederick} S.,  {Kara} E.,  {Reynolds} C.,  {Pinto} C.,   {Fabian} A.,  2018,
  \mn@doi [\apj] {10.3847/1538-4357/aae306}, \href
  {https://ui.adsabs.harvard.edu/#abs/2018ApJ...867...67F} {867, 67}

\bibitem[\protect\citeauthoryear{{Galeev}, {Rosner}  \& {Vaiana}}{{Galeev}
  et~al.}{1979}]{galeev79}
{Galeev} A.~A.,  {Rosner} R.,   {Vaiana} G.~S.,  1979, \mn@doi [\apj]
  {10.1086/156957}, \href {http://adsabs.harvard.edu/abs/1979ApJ...229..318G}
  {229, 318}

\bibitem[\protect\citeauthoryear{{Gallo}, {Miniutti}, {Miller}, {Brenneman},
  {Fabian}, {Guainazzi}  \& {Reynolds}}{{Gallo} et~al.}{2011a}]{gallo11}
{Gallo} L.~C.,  {Miniutti} G.,  {Miller} J.~M.,  {Brenneman} L.~W.,  {Fabian}
  A.~C.,  {Guainazzi} M.,   {Reynolds} C.~S.,  2011a, \mn@doi [\mnras]
  {10.1111/j.1365-2966.2010.17705.x}, \href
  {https://ui.adsabs.harvard.edu/\#abs/2011MNRAS.411..607G} {411, 607}

\bibitem[\protect\citeauthoryear{{Gallo}, {Grupe}, {Schartel}, {Komossa},
  {Miniutti}, {Fabian}  \& {Santos-Lleo}}{{Gallo} et~al.}{2011b}]{gallo11b}
{Gallo} L.~C.,  {Grupe} D.,  {Schartel} N.,  {Komossa} S.,  {Miniutti} G.,
  {Fabian} A.~C.,   {Santos-Lleo} M.,  2011b, \mn@doi [\mnras]
  {10.1111/j.1365-2966.2010.17894.x}, \href
  {https://ui.adsabs.harvard.edu/\#abs/2011MNRAS.412..161G} {412, 161}

\bibitem[\protect\citeauthoryear{{Gallo} et~al.,}{{Gallo}
  et~al.}{2015}]{gallo15}
{Gallo} L.~C.,  et~al., 2015, \mn@doi [\mnras] {10.1093/mnras/stu2108}, \href
  {https://ui.adsabs.harvard.edu/\#abs/2015MNRAS.446..633G} {446, 633}

\bibitem[\protect\citeauthoryear{{Garc{\'{\i}}a} \& {Kallman}}{{Garc{\'{\i}}a}
  \& {Kallman}}{2010}]{garcia10}
{Garc{\'{\i}}a} J.,  {Kallman} T.~R.,  2010, \mn@doi [\apj]
  {10.1088/0004-637X/718/2/695}, \href
  {http://adsabs.harvard.edu/abs/2010ApJ...718..695G} {718, 695}

\bibitem[\protect\citeauthoryear{{Garc{\'{\i}}a}, {Fabian}, {Kallman},
  {Dauser}, {Parker}, {McClintock}, {Steiner}  \& {Wilms}}{{Garc{\'{\i}}a}
  et~al.}{2016}]{garcia16}
{Garc{\'{\i}}a} J.~A.,  {Fabian} A.~C.,  {Kallman} T.~R.,  {Dauser} T.,
  {Parker} M.~L.,  {McClintock} J.~E.,  {Steiner} J.~F.,   {Wilms} J.,  2016,
  \mn@doi [\mnras] {10.1093/mnras/stw1696}, \href
  {http://adsabs.harvard.edu/abs/2016MNRAS.462..751G} {462, 751}

\bibitem[\protect\citeauthoryear{{Garcia} et~al.,}{{Garcia}
  et~al.}{2018}]{garcia18}
{Garcia} J.~A.,  et~al., 2018, arXiv e-prints, \href
  {https://ui.adsabs.harvard.edu/\#abs/2018arXiv181203194G} {p.
  arXiv:1812.03194}

\bibitem[\protect\citeauthoryear{{Grevesse}, {Noels}  \& {Sauval}}{{Grevesse}
  et~al.}{1996}]{grevesse96}
{Grevesse} N.,  {Noels} A.,   {Sauval} A.~J.,  1996, in {Holt} S.~S.,
  {Sonneborn} G.,  eds,  Astronomical Society of the Pacific Conference Series
  Vol. 99, Cosmic Abundances. p.~117

\bibitem[\protect\citeauthoryear{{Grupe}, {Komossa}  \& {Gallo}}{{Grupe}
  et~al.}{2007}]{grupe07}
{Grupe} D.,  {Komossa} S.,   {Gallo} L.~C.,  2007, \mn@doi [\apj]
  {10.1086/523042}, \href
  {https://ui.adsabs.harvard.edu/\#abs/2007ApJ...668L.111G} {668, L111}

\bibitem[\protect\citeauthoryear{{Grupe}, {Komossa}, {Gallo}, {Longinotti},
  {Fabian}, {Pradhan}, {Gruberbauer}  \& {Xu}}{{Grupe} et~al.}{2012}]{grupe12}
{Grupe} D.,  {Komossa} S.,  {Gallo} L.~C.,  {Longinotti} A.~L.,  {Fabian}
  A.~C.,  {Pradhan} A.~K.,  {Gruberbauer} M.,   {Xu} D.,  2012, \mn@doi [The
  Astrophysical Journal Supplement Series] {10.1088/0067-0049/199/2/28}, \href
  {https://ui.adsabs.harvard.edu/\#abs/2012ApJS..199...28G} {199, 28}

\bibitem[\protect\citeauthoryear{{Haardt} \& {Maraschi}}{{Haardt} \&
  {Maraschi}}{1991}]{haardt91}
{Haardt} F.,  {Maraschi} L.,  1991, \mn@doi [\apjl] {10.1086/186171}, \href
  {http://adsabs.harvard.edu/abs/1991ApJ...380L..51H} {380, L51}

\bibitem[\protect\citeauthoryear{{Jiang} et~al.,}{{Jiang}
  et~al.}{2018}]{jiang18}
{Jiang} J.,  et~al., 2018, \mn@doi [\mnras] {10.1093/mnras/sty836}, \href
  {http://adsabs.harvard.edu/abs/2018MNRAS.477.3711J} {477, 3711}

\bibitem[\protect\citeauthoryear{{Jiang}, {Walton}, {Fabian}  \&
  {Parker}}{{Jiang} et~al.}{2019a}]{jiang18d}
{Jiang} J.,  {Walton} D.~J.,  {Fabian} A.~C.,   {Parker} M.~L.,  2019a, \mn@doi
  [\mnras] {10.1093/mnras/sty3228}, \href
  {https://ui.adsabs.harvard.edu/\#abs/2019MNRAS.483.2958J} {483, 2958}

\bibitem[\protect\citeauthoryear{{Jiang}, {Fabian}, {Wang}, {Walton},
  {Garc{\'\i}a}, {Parker}, {Steiner}  \& {Tomsick}}{{Jiang}
  et~al.}{2019b}]{jiang19}
{Jiang} J.,  {Fabian} A.~C.,  {Wang} J.,  {Walton} D.~J.,  {Garc{\'\i}a} J.~A.,
   {Parker} M.~L.,  {Steiner} J.~F.,   {Tomsick} J.~A.,  2019b, \mn@doi
  [\mnras] {10.1093/mnras/stz095}, \href
  {https://ui.adsabs.harvard.edu/abs/2019MNRAS.484.1972J} {484, 1972}

\bibitem[\protect\citeauthoryear{{Kallman} \& {Bautista}}{{Kallman} \&
  {Bautista}}{2001}]{kallman01}
{Kallman} T.,  {Bautista} M.,  2001, \mn@doi [\apjs] {10.1086/319184}, \href
  {http://adsabs.harvard.edu/abs/2001ApJS..133..221K} {133, 221}

\bibitem[\protect\citeauthoryear{{Kara}, {Fabian}, {Cackett}, {Uttley},
  {Wilkins}  \& {Zoghbi}}{{Kara} et~al.}{2013}]{kara13}
{Kara} E.,  {Fabian} A.~C.,  {Cackett} E.~M.,  {Uttley} P.,  {Wilkins} D.~R.,
  {Zoghbi} A.,  2013, \mn@doi [\mnras] {10.1093/mnras/stt1055}, \href
  {http://adsabs.harvard.edu/abs/2013MNRAS.434.1129K} {434, 1129}

\bibitem[\protect\citeauthoryear{{Kara} et~al.,}{{Kara} et~al.}{2015}]{kara15}
{Kara} E.,  et~al., 2015, \mn@doi [\mnras] {10.1093/mnras/stu2136}, \href
  {https://ui.adsabs.harvard.edu/\#abs/2015MNRAS.446..737K} {446, 737}

\bibitem[\protect\citeauthoryear{{Kara}, {Alston}, {Fabian}, {Cackett},
  {Uttley}, {Reynolds}  \& {Zoghbi}}{{Kara} et~al.}{2016}]{kara16}
{Kara} E.,  {Alston} W.~N.,  {Fabian} A.~C.,  {Cackett} E.~M.,  {Uttley} P.,
  {Reynolds} C.~S.,   {Zoghbi} A.,  2016, \mn@doi [\mnras]
  {10.1093/mnras/stw1695}, \href
  {https://ui.adsabs.harvard.edu/\#abs/2016MNRAS.462..511K} {462, 511}

\bibitem[\protect\citeauthoryear{{Kara}, {Garc{\'\i}a}, {Lohfink}, {Fabian},
  {Reynolds}, {Tombesi}  \& {Wilkins}}{{Kara} et~al.}{2017}]{kara17}
{Kara} E.,  {Garc{\'\i}a} J.~A.,  {Lohfink} A.,  {Fabian} A.~C.,  {Reynolds}
  C.~S.,  {Tombesi} F.,   {Wilkins} D.~R.,  2017, \mn@doi [\mnras]
  {10.1093/mnras/stx792}, \href
  {https://ui.adsabs.harvard.edu/#abs/2017MNRAS.468.3489K} {468, 3489}

\bibitem[\protect\citeauthoryear{{Kaspi}, {Smith}, {Netzer}, {Maoz}, {Jannuzi}
  \& {Giveon}}{{Kaspi} et~al.}{2000}]{kaspi00}
{Kaspi} S.,  {Smith} P.~S.,  {Netzer} H.,  {Maoz} D.,  {Jannuzi} B.~T.,
  {Giveon} U.,  2000, \mn@doi [\apj] {10.1086/308704}, \href
  {http://adsabs.harvard.edu/abs/2000ApJ...533..631K} {533, 631}

\bibitem[\protect\citeauthoryear{{Longinotti}, {Bianchi}, {Santos-Lleo},
  {Rodr{\'\i}guez-Pascual}, {Guainazzi}, {Cardaci}  \& {Pollock}}{{Longinotti}
  et~al.}{2007}]{longinotti07}
{Longinotti} A.~L.,  {Bianchi} S.,  {Santos-Lleo} M.,  {Rodr{\'\i}guez-Pascual}
  P.,  {Guainazzi} M.,  {Cardaci} M.,   {Pollock} A.~M.~T.,  2007, \mn@doi
  [\aap] {10.1051/0004-6361:20066248}, \href
  {https://ui.adsabs.harvard.edu/\#abs/2007A&A...470...73L} {470, 73}

\bibitem[\protect\citeauthoryear{{Longinotti} et~al.,}{{Longinotti}
  et~al.}{2013}]{longinotti13}
{Longinotti} A.~L.,  et~al., 2013, \mn@doi [\apj]
  {10.1088/0004-637X/766/2/104}, \href
  {https://ui.adsabs.harvard.edu/\#abs/2013ApJ...766..104L} {766, 104}

\bibitem[\protect\citeauthoryear{{Mallick} et~al.,}{{Mallick}
  et~al.}{2018}]{mallick18}
{Mallick} L.,  et~al., 2018, \mn@doi [\mnras] {10.1093/mnras/sty1487}, \href
  {https://ui.adsabs.harvard.edu/#abs/2018MNRAS.479..615M} {479, 615}

\bibitem[\protect\citeauthoryear{{Marinucci} et~al.,}{{Marinucci}
  et~al.}{2014}]{marinucci14}
{Marinucci} A.,  et~al., 2014, \mn@doi [\mnras] {10.1093/mnras/stu404}, \href
  {https://ui.adsabs.harvard.edu/#abs/2014MNRAS.440.2347M} {440, 2347}

\bibitem[\protect\citeauthoryear{{Matt} et~al.,}{{Matt} et~al.}{2014}]{matt14}
{Matt} G.,  et~al., 2014, \mn@doi [\mnras] {10.1093/mnras/stu159}, \href
  {http://adsabs.harvard.edu/abs/2014MNRAS.439.3016M} {439, 3016}

\bibitem[\protect\citeauthoryear{{McLure} \& {Dunlop}}{{McLure} \&
  {Dunlop}}{2004}]{mclure04}
{McLure} R.~J.,  {Dunlop} J.~S.,  2004, \mn@doi [\mnras]
  {10.1111/j.1365-2966.2004.08034.x}, \href
  {http://adsabs.harvard.edu/abs/2004MNRAS.352.1390M} {352, 1390}

\bibitem[\protect\citeauthoryear{{Miniutti}, {Panessa}, {de Rosa}, {Fabian},
  {Malizia}, {Molina}, {Miller}  \& {Vaughan}}{{Miniutti}
  et~al.}{2009}]{miniutti09}
{Miniutti} G.,  {Panessa} F.,  {de Rosa} A.,  {Fabian} A.~C.,  {Malizia} A.,
  {Molina} M.,  {Miller} J.~M.,   {Vaughan} S.,  2009, \mn@doi [\mnras]
  {10.1111/j.1365-2966.2009.15092.x}, \href
  {https://ui.adsabs.harvard.edu/\#abs/2009MNRAS.398..255M} {398, 255}

\bibitem[\protect\citeauthoryear{{Nagao}, {Murayama}  \& {Taniguchi}}{{Nagao}
  et~al.}{2001}]{nagao01}
{Nagao} T.,  {Murayama} T.,   {Taniguchi} Y.,  2001, \mn@doi [\apj]
  {10.1086/318300}, \href
  {https://ui.adsabs.harvard.edu/\#abs/2001ApJ...546..744N} {546, 744}

\bibitem[\protect\citeauthoryear{{Nardini}, {Fabian}  \& {Walton}}{{Nardini}
  et~al.}{2012}]{nardini12}
{Nardini} E.,  {Fabian} A.~C.,   {Walton} D.~J.,  2012, \mn@doi [\mnras]
  {10.1111/j.1365-2966.2012.21123.x}, \href
  {https://ui.adsabs.harvard.edu/\#abs/2012MNRAS.423.3299N} {423, 3299}

\bibitem[\protect\citeauthoryear{{Nardini}, {Porquet}, {Reeves}, {Braito},
  {Lobban}  \& {Matt}}{{Nardini} et~al.}{2016}]{nardini16}
{Nardini} E.,  {Porquet} D.,  {Reeves} J.~N.,  {Braito} V.,  {Lobban} A.,
  {Matt} G.,  2016, \mn@doi [\apj] {10.3847/0004-637X/832/1/45}, \href
  {https://ui.adsabs.harvard.edu/\#abs/2016ApJ...832...45N} {832, 45}

\bibitem[\protect\citeauthoryear{{Orosz}, {McClintock}, {Aufdenberg},
  {Remillard}, {Reid}, {Narayan}  \& {Gou}}{{Orosz} et~al.}{2011}]{orosz11}
{Orosz} J.~A.,  {McClintock} J.~E.,  {Aufdenberg} J.~P.,  {Remillard} R.~A.,
  {Reid} M.~J.,  {Narayan} R.,   {Gou} L.,  2011, \mn@doi [\apj]
  {10.1088/0004-637X/742/2/84}, \href
  {https://ui.adsabs.harvard.edu/abs/2011ApJ...742...84O} {742, 84}

\bibitem[\protect\citeauthoryear{{Osterbrock} \& {Phillips}}{{Osterbrock} \&
  {Phillips}}{1977}]{osterbrock77}
{Osterbrock} D.~E.,  {Phillips} M.~M.,  1977, \mn@doi [\pasp] {10.1086/130110},
  \href {http://adsabs.harvard.edu/abs/1977PASP...89..251O} {89, 251}

\bibitem[\protect\citeauthoryear{{Page}, {Schartel}, {Turner}  \&
  {O'Brien}}{{Page} et~al.}{2004}]{page04}
{Page} K.~L.,  {Schartel} N.,  {Turner} M.~J.~L.,   {O'Brien} P.~T.,  2004,
  \mn@doi [\mnras] {10.1111/j.1365-2966.2004.07939.x}, \href
  {https://ui.adsabs.harvard.edu/\#abs/2004MNRAS.352..523P} {352, 523}

\bibitem[\protect\citeauthoryear{{Parker} et~al.,}{{Parker}
  et~al.}{2014}]{parker14}
{Parker} M.~L.,  et~al., 2014, \mn@doi [\mnras] {10.1093/mnras/stu1246}, \href
  {http://adsabs.harvard.edu/abs/2014MNRAS.443.1723P} {443, 1723}

\bibitem[\protect\citeauthoryear{{Parker} et~al.,}{{Parker}
  et~al.}{2016}]{parker16}
{Parker} M.~L.,  et~al., 2016, \mn@doi [\apjl] {10.3847/2041-8205/821/1/L6},
  \href {http://adsabs.harvard.edu/abs/2016ApJ...821L...6P} {821, L6}

\bibitem[\protect\citeauthoryear{{Parker}, {Miller}  \& {Fabian}}{{Parker}
  et~al.}{2018}]{parker18}
{Parker} M.~L.,  {Miller} J.~M.,   {Fabian} A.~C.,  2018, \mn@doi [\mnras]
  {10.1093/mnras/stx2861}, \href
  {http://adsabs.harvard.edu/abs/2018MNRAS.474.1538P} {474, 1538}

\bibitem[\protect\citeauthoryear{{Pei}}{{Pei}}{1992}]{pei92}
{Pei} Y.~C.,  1992, \mn@doi [\apj] {10.1086/171637}, \href
  {https://ui.adsabs.harvard.edu/#abs/1992ApJ...395..130P} {395, 130}

\bibitem[\protect\citeauthoryear{{Petrucci} et~al.,}{{Petrucci}
  et~al.}{2013}]{petrucci13}
{Petrucci} P.~O.,  et~al., 2013, \mn@doi [\aap] {10.1051/0004-6361/201219956},
  \href {https://ui.adsabs.harvard.edu/\#abs/2013A&A...549A..73P} {549, A73}

\bibitem[\protect\citeauthoryear{{Porquet} et~al.,}{{Porquet}
  et~al.}{2019}]{porquet19}
{Porquet} D.,  et~al., 2019, arXiv e-prints, \href
  {https://ui.adsabs.harvard.edu/\#abs/2019arXiv190101812P} {p.
  arXiv:1901.01812}

\bibitem[\protect\citeauthoryear{{Raimundo}, {Fabian}, {Vasudevan}, {Gandhi}
  \& {Wu}}{{Raimundo} et~al.}{2012}]{raimundo12}
{Raimundo} S.~I.,  {Fabian} A.~C.,  {Vasudevan} R.~V.,  {Gandhi} P.,   {Wu} J.,
   2012, \mn@doi [\mnras] {10.1111/j.1365-2966.2011.19904.x}, \href
  {https://ui.adsabs.harvard.edu/\#abs/2012MNRAS.419.2529R} {419, 2529}

\bibitem[\protect\citeauthoryear{{Reeves}, {Porquet}, {Braito}, {Nardini},
  {Lobban}  \& {Turner}}{{Reeves} et~al.}{2016}]{reeves16}
{Reeves} J.~N.,  {Porquet} D.,  {Braito} V.,  {Nardini} E.,  {Lobban} A.,
  {Turner} T.~J.,  2016, \mn@doi [\apj] {10.3847/0004-637X/828/2/98}, \href
  {https://ui.adsabs.harvard.edu/\#abs/2016ApJ...828...98R} {828, 98}

\bibitem[\protect\citeauthoryear{{Reid}, {McClintock}, {Narayan}, {Gou},
  {Remillard}  \& {Orosz}}{{Reid} et~al.}{2011}]{reid11}
{Reid} M.~J.,  {McClintock} J.~E.,  {Narayan} R.,  {Gou} L.,  {Remillard}
  R.~A.,   {Orosz} J.~A.,  2011, \mn@doi [\apj] {10.1088/0004-637X/742/2/83},
  \href {https://ui.adsabs.harvard.edu/abs/2011ApJ...742...83R} {742, 83}

\bibitem[\protect\citeauthoryear{{Risaliti}, {Maiolino}  \&
  {Salvati}}{{Risaliti} et~al.}{1999}]{risaliti99}
{Risaliti} G.,  {Maiolino} R.,   {Salvati} M.,  1999, \mn@doi [\apj]
  {10.1086/307623}, \href {http://adsabs.harvard.edu/abs/1999ApJ...522..157R}
  {522, 157}

\bibitem[\protect\citeauthoryear{{Ross} \& {Fabian}}{{Ross} \&
  {Fabian}}{1993}]{ross93}
{Ross} R.~R.,  {Fabian} A.~C.,  1993, \mn@doi [\mnras]
  {10.1093/mnras/261.1.74}, \href
  {https://ui.adsabs.harvard.edu/\#abs/1993MNRAS.261...74R} {261, 74}

\bibitem[\protect\citeauthoryear{{Ross} \& {Fabian}}{{Ross} \&
  {Fabian}}{2007}]{ross07}
{Ross} R.~R.,  {Fabian} A.~C.,  2007, \mn@doi [\mnras]
  {10.1111/j.1365-2966.2007.12339.x}, \href
  {http://adsabs.harvard.edu/abs/2007MNRAS.381.1697R} {381, 1697}

\bibitem[\protect\citeauthoryear{{Shakura} \& {Sunyaev}}{{Shakura} \&
  {Sunyaev}}{1973}]{shakura73}
{Shakura} N.~I.,  {Sunyaev} R.~A.,  1973, \aap, \href
  {http://cdsads.u-strasbg.fr/abs/1973A%26A....24..337S} {24, 337}

\bibitem[\protect\citeauthoryear{{Stella} \& {Rosner}}{{Stella} \&
  {Rosner}}{1984}]{stella84}
{Stella} L.,  {Rosner} R.,  1984, \mn@doi [\apj] {10.1086/161697}, \href
  {http://adsabs.harvard.edu/abs/1984ApJ...277..312S} {277, 312}

\bibitem[\protect\citeauthoryear{{Svensson} \& {Zdziarski}}{{Svensson} \&
  {Zdziarski}}{1994}]{svensson94}
{Svensson} R.,  {Zdziarski} A.~A.,  1994, \mn@doi [\apj] {10.1086/174934},
  \href {http://cdsads.u-strasbg.fr/abs/1994ApJ...436..599S} {436, 599}

\bibitem[\protect\citeauthoryear{{Takahashi}, {Hayashida}  \&
  {Anabuki}}{{Takahashi} et~al.}{2010}]{takahashi10}
{Takahashi} H.,  {Hayashida} K.,   {Anabuki} N.,  2010, \mn@doi [Publications
  of the Astronomical Society of Japan] {10.1093/pasj/62.6.1483}, \href
  {https://ui.adsabs.harvard.edu/\#abs/2010PASJ...62.1483T} {62, 1483}

\bibitem[\protect\citeauthoryear{{Tomsick} et~al.,}{{Tomsick}
  et~al.}{2018}]{tomsick18}
{Tomsick} J.~A.,  et~al., 2018, \mn@doi [\apj] {10.3847/1538-4357/aaaab1},
  \href {http://adsabs.harvard.edu/abs/2018ApJ...855....3T} {855, 3}

\bibitem[\protect\citeauthoryear{{Vasudevan} \& {Fabian}}{{Vasudevan} \&
  {Fabian}}{2007}]{vasudevan07}
{Vasudevan} R.~V.,  {Fabian} A.~C.,  2007, \mn@doi [\mnras]
  {10.1111/j.1365-2966.2007.12328.x}, \href
  {http://adsabs.harvard.edu/abs/2007MNRAS.381.1235V} {381, 1235}

\bibitem[\protect\citeauthoryear{{Vaughan}, {Boller}, {Fabian}, {Ballantyne},
  {Brandt}  \& {Tr{\"u}mper}}{{Vaughan} et~al.}{2002}]{vaughan02}
{Vaughan} S.,  {Boller} T.,  {Fabian} A.~C.,  {Ballantyne} D.~R.,  {Brandt}
  W.~N.,   {Tr{\"u}mper} J.,  2002, \mn@doi [\mnras]
  {10.1046/j.1365-8711.2002.05908.x}, \href
  {https://ui.adsabs.harvard.edu/\#abs/2002MNRAS.337..247V} {337, 247}

\bibitem[\protect\citeauthoryear{{V{\'e}ron-Cetty} \&
  {V{\'e}ron}}{{V{\'e}ron-Cetty} \& {V{\'e}ron}}{2006}]{veron06}
{V{\'e}ron-Cetty} M.~P.,  {V{\'e}ron} P.,  2006, \mn@doi [\aap]
  {10.1051/0004-6361:20065177}, \href
  {https://ui.adsabs.harvard.edu/\#abs/2006A&A...455..773V} {455, 773}

\bibitem[\protect\citeauthoryear{{Walton}, {Reis}, {Cackett}, {Fabian}  \&
  {Miller}}{{Walton} et~al.}{2012}]{walton12}
{Walton} D.~J.,  {Reis} R.~C.,  {Cackett} E.~M.,  {Fabian} A.~C.,   {Miller}
  J.~M.,  2012, \mn@doi [\mnras] {10.1111/j.1365-2966.2012.20809.x}, \href
  {http://adsabs.harvard.edu/abs/2012MNRAS.422.2510W} {422, 2510}

\bibitem[\protect\citeauthoryear{{Walton}, {Nardini}, {Fabian}, {Gallo}  \&
  {Reis}}{{Walton} et~al.}{2013}]{walton13}
{Walton} D.~J.,  {Nardini} E.,  {Fabian} A.~C.,  {Gallo} L.~C.,   {Reis} R.~C.,
   2013, \mn@doi [\mnras] {10.1093/mnras/sts227}, \href
  {https://ui.adsabs.harvard.edu/\#abs/2013MNRAS.428.2901W} {428, 2901}

\bibitem[\protect\citeauthoryear{{Wang-Ji} et~al.,}{{Wang-Ji}
  et~al.}{2018}]{jingyi18}
{Wang-Ji} J.,  et~al., 2018, \mn@doi [\apj] {10.3847/1538-4357/aaa974}, \href
  {http://adsabs.harvard.edu/abs/2018ApJ...855...61W} {855, 61}

\bibitem[\protect\citeauthoryear{{Willingale}, {Starling}, {Beardmore},
  {Tanvir}  \& {O'Brien}}{{Willingale} et~al.}{2013}]{willingale13}
{Willingale} R.,  {Starling} R.~L.~C.,  {Beardmore} A.~P.,  {Tanvir} N.~R.,
  {O'Brien} P.~T.,  2013, \mn@doi [\mnras] {10.1093/mnras/stt175}, \href
  {http://adsabs.harvard.edu/abs/2013MNRAS.431..394W} {431, 394}

\bibitem[\protect\citeauthoryear{{Zoghbi}, {Fabian}, {Reynolds}  \&
  {Cackett}}{{Zoghbi} et~al.}{2012}]{zoghbi12}
{Zoghbi} A.,  {Fabian} A.~C.,  {Reynolds} C.~S.,   {Cackett} E.~M.,  2012,
  \mn@doi [\mnras] {10.1111/j.1365-2966.2012.20587.x}, \href
  {https://ui.adsabs.harvard.edu/\#abs/2012MNRAS.422..129Z} {422, 129}

\makeatother
\end{thebibliography}

%%%%%%%%%%%%%%%%%%%%%%%%%%%%%%%%%%%%%%%%%%%%%%%%%%

%%%%%%%%%%%%%%%%% APPENDICES %%%%%%%%%%%%%%%%%%%%%

\appendix

\section{Observation and Spectral Analysis Details} \label{re1}

\subsection{1H~1934$-$603}
1H~1934$-$603 is a narrow-line Seyfert 1 galaxy \citep[NLS1,][]{nagao01} that shows fast variability in the X-ray band. Previously by analysing the archival \xmm\ and \nustar\ observations of 1H~1934$-$603, \citet{frederick18} discovered that the disc reflection component lags behind the coronal power-law continuum by $\approx20$\,s. By conducting a novel spectral analysis using a fixed disc density reflection model, a super solar iron abundance ($Z_{\rm Fe}>9Z_{\odot}$) is required for the reflection spectral modelling \citep{frederick18}.

A ratio plot for an averaged EPIC-pn spectrum of 1H~1934$-$603 against an absorbed  power-law model is shown in Figure \ref{pic_fe}. A broad emission line feature is visible in the iron band. By fitting the emission line with a simple Gaussian line model \texttt{zgauss}, we obtain a best-fit rest-frame line energy at $E_{\rm line}=6.65\pm0.05$\,keV with $\sigma=0.49^{+0.08}_{-0.07}$\,keV. The equivalent width (EW) of the emission line is $264^{+8}_{-7}$\,eV. Small residuals are visible at 5~keV when the line feature is modelled by a simple Gaussian line model, requiring a more physical modelling for the broad emission line (e.g. relativistic disc line). No obvious narrow line component has been found. 

Based on the spectral analysis in the iron band, we then model the broad band spectrum with \texttt{MODEL2}. \texttt{MODEL2} can provide a very good fit with C-Stat/$\nu$=242.35/180. The best-fit model and corresponding ratio plot are shown in Figure \ref{pic_final1}. No structural residuals are found in the ratio plot. A disc density of $\log(n_{\rm e})=17.7^{+0.2}_{-0.3}$ is required with a disc iron abundance of $Z_{\rm Fe}/Z_{\odot}=5.9^{+0.6}_{-1.4}$. The iron abundance obtained with the disc density as a free parameter is much lower than the value obtained in previous analysis \citep{frederick18}. We obtain a lower limit of the BH spin $a_*>0.4$, which is higher than the previous analysis \citep[$a_*<0.1$,][]{frederick18}. Note that the previous analysis in \citet{frederick18} models the soft excess emission with an additional blackbody model and assumes a fixed disc density at $\log(n_{\rm e})=15$. In this work, we model both the soft excess emission and the broad iron emission line with only one disc reflection model by allowing the disc density to be a free parameter.

\subsection{Ark~120}

Ark~120 is a nearby Seyert 1 galaxy \citep[e.g.][]{osterbrock77} that is well-studied in the X-ray band. This source shows little or no evidence for X-ray absorption \citep[e.g.][]{reeves16}. Previous spectral analysis of Ark~120 shows evidence for three line components in the iron band \citep{nardini16}. Two of the three line components are narrow emission lines, corresponding to a neutral Fe~K$\alpha$ emission line and an ionized Fe~K$\alpha$ emission line. The third line component is broader with FWHM$\approx5000$\,km\,s$^{-1}$.

A ratio plot for an averaged EPIC-pn spectrum of Ark~120 against an absorbed power-law model is shown in Figure \ref{pic_fe}. A combination of narrow emission lines and a broad emission line is shown in the iron band. The line shapes are similar to these found in \citet{nardini16}. By fitting the line features with three Gaussian line model \texttt{zgauss}, we obtain two of the three line components are at $6.43^{+0.05}_{-0.02}$\,keV (EW=$38^{+42}_{-12}$\,eV) and $7.03^{+0.03}_{-0.02}$\,keV (EW=$27^{+20}_{-12}$\,eV). The best-fit line widths for these two line components are $<0.01$\,keV and $0.06^{+0.04}_{-0.05}$\,keV correspondingly. The 6.43~keV emission line can be interpreted as the neutral Fe~K$\alpha$ emission line and the other line can be interpreted as the hydrogenic iron. The third line component is located at $6.49^{+0.05}_{-0.02}$\,keV (EW=$100^{+14}_{-13}$). The width of the line is $0.30^{+0.06}_{-0.05}$\,keV, indicating a broad emission line from the inner disc region.

Based on the existence of both a neutral and ionized narrow iron emission lines, we model the broad band spectrum with  \texttt{MODEL3}. \texttt{MODEL3} can provide a good fit with C-stat/$\nu$=288.61/168. The best-fit model and corresponding ratio plot are shown in Figure \ref{pic_final1}. Only an upper limit of the disc density is found $\log(n_{\rm e})<15.6$. A solar iron abundance is required for the spectral modelling. A fixed spin $a_*=0$ and a fixed viewing angle $i=30^\circ$ are assumed in \citet{nardini16}. In contrast, we obtain a high black hole spin $a_*>0.85$ and a high viewing angle $i=67^{+4}_{-5}$$^{\circ}$, which are consistent with previous reflection-based analysis of \suzaku\ observations of the same source \citep[e.g. $a_*\approx0.81$, $i\approx54^{\circ}$,][]{walton13} and other spin measurement methods \citep[e.g.][]{porquet19}.

\subsection{Ark~564}

Ark~564 is a very variable NLS1 in the X-ray band. Detailed studies of its X-ray reverberation lags with \xmm\ observations have been done in previous analyses \citep[e.g.][]{kara13}. A high iron abundance ($Z_{\rm Fe}/Z_{\odot}\approx3$) was obtained by analysing the simultaneous \suzaku\ and \nustar\ spectra above 1~keV \citep{kara17}. 

We present a broad band spectral analysis of the averaged EPIC-pn spectrum of Ark~564 with a total net pn exposure of 402~ks. A ratio plot against an absorbed power-law model is shown in Figure \ref{pic_fe}. A very strong emission line feature is shown in the iron band and a very strong soft excess is shown below 3~keV. Fitting the emission line in the iron band with \texttt{zgauss} offers a good fit with some remaining residuals at 5.5~keV, requiring more physical modelling (e.g. relativistic disc reflection model). The central energy of the line is at $E_{\rm line}=6.59^{+0.06}_{-0.07}$\,keV in the source frame with a line width of $\sigma=0.44^{+0.09}_{-0.08}$\,keV. The equivalent width of the best-fit line model is $120^{+12}_{-8}$\,eV. No obvious narrow emission line feature at 6.4~keV is found in the iron band. 

We fit the full band spectrum with \texttt{MODEL2} due to the lack of evidence for narrow emission lines in the iron band. \texttt{MODEL2} offers a very good fit for the averaged EPIC-pn spectrum of Ark~564 with C-stat/$\nu$=191.32/180. The best-fit model and corresponding ratio plot are shown in Figure \ref{pic_final1}. A close-to-solar iron abundance is obtained and a very high disc density of $\log(n_{\rm e})=18.55\pm0.07$ is required for the spectral fitting. No additional component is required to model the soft excess. A high BH spin of $a_*>0.9$ is found, similar to the previous analysis by analysing \suzaku\ observations of the same source \citep[$a_*\approx0.96$,][]{walton13}. The BH spin parameter was however not constrained in \citet{kara17}. An inclination angle of $i=57^{+4}_{-3}$$^{\circ}$ is obtained, which is consistent with the results in \citet{walton13} and \citet{kara17}.

\subsection{Mrk~110}

Mrk~110 is a NLS1 \citep{veron06} and has been observed by \xmm\ once for a net pn exposure of 33~ks. \citet{boller07} shows a complete analysis of the RGS and EPIC spectra. Only a narrow Fe~K emission line was found previously. 

By fitting the EPIC-pn spectrum of the only \xmm\ observation of Mrk~110, we confirm that only a narrow emission line is shown in the iron band. A ratio plot of the EPIC-pn spectrum fitted with an absorbed power-law model is shown in Fig.\,\ref{pic_fe}. By using a simple Gaussian line model \texttt{zgauss}, we obtain the best-fit line width of $\sigma<0.127$\,keV and the best-fit line energy of $E_{\rm line}=6.44\pm0.04$\,keV for this narrow emission line. The equivalent width of the line component is $51^{+13}_{-22}$\,eV. The narrow emission line is at 6.44~keV and can be interpreted as the neutral Fe~K$\alpha$ emission line.

Based on the narrow neutral Fe~K$\alpha$ emission line in the iron band, we model the broad band spectrum with \texttt{MODEL1}. \texttt{MODEL1} offers a very good fit with C-stat/$\nu$=205.35/161. The best-fit model and corresponding ratio plot are shown in Fig.\,\ref{pic_final1}. The relativistic disc reflection model accounts for mainly the soft excess below 2~keV. The lack of the broad Fe~K$\alpha$ emission line in the iron band might be due to the extremely blurred reflection component, as seen in our modelling. Only an upper limit of the disc density ($\log(n_{\rm e})<16.5$) is found.

\subsection{Mrk~1310}

Mrk~1310 is a Seyfert 1 galaxy \citep{veron06} and has only one \xmm\ observation with a net pn exposure of 35\,ks. The iron band does not show strong evidence for emission features. A power-law model can offer a very good fit for the spectra between 3--10~keV with C-stat/$\nu$=52.02/46. By adding an additional line model \texttt{zgauss} with the line energy fixed at 6.4~keV, the fit can be improved by $\Delta$C-stat=4 with 2 more free parameters. The equivalent width of the line component is $<20$\,eV. In the soft band, Mrk~1310 however shows a strong soft excess, as in other AGN in our sample.

We model the broad band spectrum of Mrk~1310 with \texttt{MODEL2}. \texttt{MODEL2} offers a very good fit with C-stat/$\nu=109.90/94$. The best-fit model and corresponding ratio plot are shown in Figure \ref{pic_final1}. The high density disc reflection component accounts for the soft excess emission. However, due to the lack of a broad Fe~K$\alpha$ emission line, we are unable to constrain the disc emissivity profile and the spin of the BH. We assume the emissivity index in a flat spacetime ($q_{1}=q_{2}=3$) and a maximum spin parameter. The best-fit parameters are shown in \red{Table\,\ref{tab_fit_info_high_ne}}. By modelling the soft excess emission with high density disc reflection model, a high disc density of $\log(n_{\rm e})=17^{+0.4}_{-0.2}$ is required.

\subsection{Mrk~279}

Previous analysis of the long \xmm\ observations of the Seyfert 1 galaxy Mrk~279 \citep{veron06} in 2005 by \citet{costantini10} shows very complex emission features in the iron band, indicating both a broad Fe~K$\alpha$ emission line from the disc and a narrow Fe~\textsc{xxvi} line potentially from the outer layer of the torus.

A ratio plot of the stacked pn spectrum of Mrk~279 fitted with an absorbed power law is shown in Fig.\,\ref{pic_fe}. The iron band of the spectrum shows two narrow emission features and  a broad line component. By modelling the line features with multiple \texttt{zgauss} models, we obtained a very good fit in the 3--10~keV band. Three \texttt{zgauss} models are required: a broad line at $6.6^{+0.5}_{-0.3}$\,keV (EW=$50^{+22}_{-13}$\,eV, $\sigma=0.38^{+0.12}_{-0.08}$eV); a narrow line at $6.41^{+0.02}_{-0.04}$\,keV (EW=$85^{+14}_{-12}$\,eV, $\sigma<0.02$eV); a second narrow line at $6.98^{+0.06}_{-0.12}$\,keV (EW=$<22$\,eV, $\sigma<0.02$eV). The second narrow line at 6.98~keV is consistent with previous analysis by \citet{costantini10}. However, the line feature is too weak to be constrained with an unconstrained equivalent width and a small statistical improvement when the line model is added to the fit ($\Delta$C-stat=3 with three more parameters).

Based on the indication of the iron band, we model the broad band spectrum of Mrk~279 with \texttt{MODEL1}. \texttt{MODEL1} offers a very good fit with C-stat/$\nu=155.59/177$. The best-fit model is shown in Fig.\,\ref{pic_final2}. The relativistic disc reflection model accounts for both the soft excess and the broad Fe~K$\alpha$ emission line in Mrk~279 A high BH spin of $a_*>0.95$ is required for the spectral modelling and only an upper limit of the disc density $\log(n_{\rm e})<16.9$ is achieved. We obtain a disc viewing angle of $i=37^{+8}_{-17}$$^{\circ}$, which is consistent with previous analysis \citep[$i<30^{\circ}$,][]{costantini10}.

\subsection{Mrk~335}

Mrk~335 is a NLS1 \citep{veron06} that has been well studied in the X-ray band. This source experienced several extremely low flux states in history \citep[e.g.][]{grupe07,parker14}. \citet{grupe12} found that the complex spectral variability can be explained by a variable disc reflection component. \citet{parker14} and \citet{gallo15} explain the low flux state spectrum of Mrk~335 with a reflection-dominated emission from the inner disc region. The spectral variability is due to strong light-bending effects in the vicinity of the central BH. The strongest supporting evidence for the reflection interpretation of the spectrum of Mrk~335 is the discovery of the reverberation lag between the reflected disc photons and the coronal continuum photons \citep{kara13}.

We first fit the stacked spectrum of Mrk~335 with an  absorbed power-law model and the ratio plot is shown in Fig.\,\ref{pic_fe}. The ratio plot shows a strong broad Fe~K$\alpha$ emission line feature and a strong soft excess below 2~keV. The result is similar to \citet{parker14}. By following the indication in \citet{parker14}, we model the broad band spectrum with \texttt{MODEL4}. \texttt{MODEL4} offers a very good fit with C-stat/$\nu=251.39/169$. The best-fit model is shown in Fig.\,\ref{pic_final2}. One thin warm absorber modelled by \texttt{warmabs} with $N_{\rm H}=2.45^{+0.39}_{-0.17}\times10^{20}$\,cm$^{-2}$ and $\log(\xi)=1.38\pm0.02$ is found. We obtain a very high reflection fraction $f_{\rm refl}=3.3^{+0.4}_{-0.6}$, indicating a reflection-dominated scenario, similar with previous analysis \citep[e.g.][]{parker14}. By fitting the broad Fe~K$\alpha$ emission line and the soft excess with the same reflection model, we obtain a very steep disc emissivity profile (see \red{Table\,\ref{tab_fit_info_high_ne}} for best-fit parameters). The large inner emissivity index $q_{1}$ and the low broken radius $R_{\rm r}$ indicate a very compact coronal region \citep[e.g. <3$r_{\rm g}$,][]{parker14}. Only an upper limit of the disc density $\log(n_{\rm e})<16.0$ is achieved. We find a very BH spin of $a_*>0.988$ and a small inclination angle of $i=33^{+4}_{-5}$, which are consistent with previous analysis of \nustar\ observations of the same source\citep[e.g. $a_*\approx0.99$, $i\approx25^{\circ}$,][]{parker14}. Our best-fit inclination angle is however lower than the value measured using \suzaku\ observations \citep[e.g. $i\approx50-58^{\circ}$,][]{walton13,gallo15}.

\subsection{Mrk~590}

Mrk~590 is a Seyfert 1 galaxy \citep{veron06}. Previous analysis of the quasi-simultaneous \xmm\ and \chandra\ observations in 2004 shows evidence for a strong soft excess and narrow Fe~K$\alpha$, Fe~\textsc{xxv} and Fe~\textsc{xxvi} emissions \citep{longinotti07}. 

We first fit the 3--10~keV band spectrum with an  absorbed power-law model and the ratio plot is shown in Fig.\ref{pic_fe}. By fitting the narrow lines features with simple \texttt{zgauss} models, Two narrow line models are required, one line at $6.407\pm0.02$\,keV (EW=$135^{+12}_{-23}$\,eV, $\sigma<0.06$\,keV) and the other line at $7.04^{+0.06}_{-0.10}$\,keV (EW=$46^{+23}_{-35}$, $\sigma<0.12$\,keV). The former line can be interpreted as Fe~K$\alpha$ emission line and the latter can be interpreted as Fe~\textsc{xxvi} line. We do not find strong evidence for a narrow Fe~\textsc{xxv} emission line as in \citet{longinotti07} or a broad emission feature.

Based on the analysis of the iron band, we model the broad band spectrum of Mrk~590 with \texttt{MODEL1}. \texttt{MODEL1} can provide a very good fit with C-stat/$\nu$=187.41/174. The best-fit model is shown in Fig.\,\ref{pic_final2}. The disc reflection component with a very high disc density of $\log(n_{\rm e})=18.4^{+0.2}_{-1.1}$ accounts for the soft excess, as shown in Fig.\,\ref{pic_final2}. We obtain a lower limit of the BH spin of $a_*>0.1$ and a high inclination angle of $i=79^{+7}_{-4}$$^{\circ}$.

\subsection{Mrk~79}

Mrk~79 is a Seyfert 1 galaxy \citep{veron06} that shows very large X-ray flux variability by a factor of 10 in multi-epoch observations \citep{gallo11b}. The spectrum of Mrk~79 shows a variable soft excess and strong narrow emission line features in the iron band. 

We stack all the EPIC-pn spectra in the archive and find two narrow emission lines in the iron band. The first line is at $6.40\pm0.02$\,keV \red{in the source frame} (EW=$173^{+32}_{-14}$\,eV, $\sigma=0.086^{+0.019}_{-0.020}$\,keV) and the second line is at $6.90\pm0.05$\,keV \red{in the source frame} (EW=$30^{+22}_{-13}$\,eV, $\sigma<0.08$\,keV), corresponding to Fe~K$\alpha$ and Fe~K$\beta$ line. 

Based on the analysis of the iron band, we model the broad band spectrum of Mrk~79 with \texttt{MODEL1}. \texttt{MODEL1} offers a very good fit with C-stat/$\nu$=225.60/177. The best-fit model is shown in Fig.\,\ref{pic_final2}. Mrk~79 has the hardest continuum emission in our sample with $\Gamma=1.712^{+0.014}_{-0.012}$. According to our calculation of the mass accretion rate using B band flux, Mrk~79 indeed has the lowest accretion rate in our sample ($\dot{m}=0.13^{+0.05}_{-0.07}$). A high disc density of $\log(n_{\rm e})=18.01^{+0.12}_{-0.32}$ is required by our spectral modelling. We obtain a BH spin of $a_*>0.5$ and an inclination angle of $i=21\pm6^{\circ}$, which is consistent with previous analyses \cite[$a_*\approx0.7$, $i\approx21^{\circ}$,][]{gallo11}.

\subsection{NGC~4748}

NGC~4748 is a NLS1 \citep{veron06} that is not well studied in the X-ray band. Only one \xmm\ observation with a net pn exposure of 26\,ks is available in the archive. 

The spectrum of NGC~4748 in the iron band shows a broad emission feature. See Fig.\,\ref{pic_fe} for a ratio plot of the spectrum of NGC~4748 fitted by an  absorbed power-law model. By fitting the line model with one \texttt{zgauss} model, the line width of the emission line is $\sigma=0.7^{+0.6}_{-0.3}$\,keV and the line is at $6.7^{+0.3}_{-0.2}$\,keV. The equivalent width of the line is $373^{+42}_{-17}$\,eV. This strong, broad 6.7\,keV emission line can be interpreted as the relativistic disc Fe~K emission line. A simple \texttt{zgauss} modelling of the line feature leaves some residuals between 7--8\,keV. No significant evidence for a narrow core is found.

Based on the evidence for a broad iron emission line feature and a soft excess, we model the broad band spectrum with \texttt{MODEL2}. \texttt{MODEL2} offers a very good fit with C-stat/$\nu$=152.66/144. Fig.\,\ref{pic_final2} shows the best-fit model and the corresponding ratio plot. One relativistic reflection model is able to model both the soft excess mission and the broad iron line feature. An intermediate disc density is required $\log(n_{\rm e})=16.6^{+0.4}_{-0.5}$. A high BH spin of $a_*>0.8$ is preferred by our model. 

\subsection{PG~0804$+$761}

PG~0804$+$761 is a Seyfert 1 galaxy \citep{veron06} and has three \xmm\ observations in the archive. The first observation in 2000 (obsID 0102040401) was entirely dominated by flaring particle background. The other two observations were taken in 2010 and have a total net pn exposure of 32~ks. By fitting the stacked pn spectrum of PG~0804$+$761 with an  absorbed power-law model, two strong emission lines are shown in the iron band. A strong soft excess is found below 2~keV. See Fig.\,\ref{pic_fe} for the ratio plot. We first model the line features with simple \texttt{zgauss} models. The line width of both emission line is $\sigma<0.08$\,keV, indicating two narrow lines from distant reflector. The first line is at $6.44\pm0.04$\,keV (EW=$99^{+35}_{-17}$\,eV) and the second line is at $6.88^{+0.06}_{-0.07}$\,keV (EW=$91^{+13}_{-24}$\,eV). These two lines can be interpreted as Fe~K$\alpha$ and Fe~K$\beta$ lines. The spectrum shows no evidence for a broad line component.

Based on the evidence for only narrow Fe~K emission lines, we model the broad band spectrum with \texttt{MODEL1}. \texttt{MODEL1} can offer a very good fit with C-stat/$\nu$=226.26/173. The relativistic disc reflection model accounts for the soft excess. See Fig.\,\ref{pic_final2} for the best-fit model and corresponding ratio plot. Only an upper limit of the disc density $\log(n_{\rm e})<15.8$ is found. 

\subsection{PG~0844$+$349}

PG~0844$+$349 is a Seyfert 1 galaxy \citep{veron06} that has shown a large flux variability of a factor of 10 in history. \citet{gallo11b} analysed the \xmm\ observation taken during the X-ray weak state of PG~0844$+$349 and found the spectrum is dominated by the disc reflection component, indicating strong light-bending effects. 

We first fit the spectrum with an  absorbed power-law model and the ratio plot is shown in Fig.\,\ref{pic_fe}. By fitting the emission features with two simple \texttt{zgauss} models, we obtain a broad line component at $6.6\pm0.2$\,keV (EW=$349^{+32}_{-24}$\,eV, $\sigma=0.31^{+0.26}_{-0.13}$\,keV) and a narrow core at 6.4\,keV (EW<42\,eV, $\sigma<0.05$\,keV). Both two line components are consistent with the results in \citet{gallo11b}. However only an upper limit of the equivalent width of the second line is obtained due to a short net exposure of only 18~ks. 

Based on the analysis in the iron band, we model the broad band spectrum with \texttt{MODEL1}. \texttt{MODEL1} can offer a very good fit with C-stat/$\nu$=163.0/129. Fig.\,\ref{pic_final3} presents the best-fit \texttt{MODEL1} for PG~0844$+$349. A disc reflection component with a high disc density parameter of $\log(n_{\rm e})=17.2^{+0.06}_{-0.27}$ can account for both the broad iron line and the soft excess emission. We obtain a high BH spin of $a_*>0.95$ and an inclination angle of $i=32^{+12}_{-15}$$^{\circ}$. The inclination angle is consistent with the measurement in \citet[][$i\approx34^{\circ}$]{gallo11b}.

\subsection{PG~1229$+$204}

PG~1229$+$204 is a Seyfert 1 galaxy \citep{veron06} and has only one \xmm\ observation with a net pn exposure of only 17~ks. Fig.\,\ref{pic_fe} shows a ratio plot of the EPIC-pn spectrum of PG~1229$+$204 against an  absorbed power-law model. The spectrum shows a strong soft excess below 2~keV and an emission feature in the iron band. By fitting the emission line with one \texttt{zgauss} model, we find that the line central energy is $6.59^{+0.15}_{-0.14}$\,keV and the line width is $\sigma=0.23^{+0.22}_{-0.11}$\,keV. The line feature is very strong with an equivalent width of EW=$209^{+23}_{-14}$\,eV. The short exposure of this observation does not allow us to distinguish a mildly broad emission feature or a combination of several emission lines (e.g. Fe~K$\alpha$, Fe~K$\beta$).

Because of the uncertainty of the nature of the emission feature in the iron band, we first fit the broad band spectrum with \texttt{MODEL2} to check if the relativistic disc reflection model is able to model both the soft excess emission and the emission feature in the iron band. \texttt{MODEL2} can offer a good fit with C-stat/$\nu$=122.23/129. However there are still residuals in the iron band indicating that the emission feature can not be modelled with a relativistic disc reflection model. Second, we fit the spectrum with \texttt{MODEL1}, including a distant neutral reflector. \texttt{MODEL1} is able to improve the fit by $\Delta$C-stat=11 with one more parameter with no structural residuals in the iron band. The best-fit model and the corresponding ratio plot are shown in Fig.\,\ref{pic_final3}. The relativistic disc reflection component with a modest disc density of $\log(n_{\rm e})=16.8^{+0.2}_{-0.3}$ accounts for the soft excess emission. We obtain a high BH spin of $a_*=0.93^{+0.06}_{-0.02}$ and an inclination angle of $i=22^{+18}_{-10}$$^{\circ}$.

\subsection{PG~1426$+$015}

PG~1426$+$015 is a Seyfert 1 galaxy \citep{veron06} and shows a blackbody-like soft excess emission \citep[e.g. kT$\approx0.1$\,keV,][]{page04}. Fig.\,\ref{pic_fe} presents a ratio plot of PG~1426$+$015 EPIC-pn spectrum fitted with an  absorbed power-law model. The spectrum shows evidence for weak emission feature in the iron band and a strong soft excess emission. By fitting the emission feature of a \texttt{zgauss} model with a fixed line energy at 6.4\,keV, we obtain an upper limit of the line width $\sigma<0.40$\,keV. Although the spectrum shows tentative evidence for a broad emission feature, the short exposure of the observation does not allow us to better constrain the line shape.

We model the broad band spectrum of PG~1426$+$015 with \texttt{MODEL2}. \texttt{MODEL2} offers a very good fit with C-stat/$\nu$=53.43/52. The best-fit model is shown in Fig.\,\ref{pic_final3}. The emissivity profile, the BH spin and the viewing angle are not constrained during our fit due to the short exposure. Therefore we fix the emissivity index at $q_{1}=q_{2}=3$, a maximum BH spin, and a viewing angle of $i=30^\circ$. We only obtain an upper limit of the disc density $\log(n_{\rm e})<15.9$.

\subsection{Swift~J2127.4$+$5654}

Swift~J2127.4$+$5654 is a Seyfert 1 galaxy \citep{veron06} that has been well studied in the X-ray band. \citet{miniutti09} analysed the \suzaku\ observations of this source and obtained a BH spin of $a_*=0.6\pm0.2$. This result has been confirmed by \citet{marinucci14} as well where \xmm\ observations are considered. A combination of broad Fe~K$\alpha$ emission line and a narrow core is found in the iron band. The \suzaku\ spectrum of Swift~J2127.4$+$5654 shows a blackbody-shaped soft excess \citet{miniutti09}. \citet{kara15} analysed the \nustar\ and found a reverberation lag of both the iron line and the Compton hump, supporting the disc reflection interpretation of the broad band spectrum of this source.

By following the indication of \citet{marinucci14}, we model the broad band spectrum with \texttt{MODEL4}. The best-fit model is shown in Fig.\,\ref{pic_final3}. An additional neutral absorber is required and the model \texttt{ztbabs} is used for this purpose. The redshift parameter of \texttt{ztbabs} is fixed at the value of the source. The best-fit model is shown in Fig.\,\ref{pic_final3}. \texttt{MODEL4} offers a good fit with C-stat/$\nu$=246.21/182. An intermediate spin of $a_*=0.72^{+0.14}_{-0.20}$ is found and is consistent with previous spin measurements \citep[e.g.][]{miniutti09,marinucci14}. A higher inclination angle ($i=67^{+3}_{-2}$$^\circ$) is found by fitting with a high density disc reflection model compared with previous analysis \citep[e.g. 40$^\circ$,][]{miniutti09}. A lower limit of the disc density ($\log(n_{\rm e})>18.7$) is obtained, indicating a potential high disc density. 

\subsection{Ton~S180}

Ton~S180 is a NLS1 galaxy \citep{veron06} that shows both a broad Fe~K emission line and a strong soft excess \citep[e.g.][]{vaughan02, takahashi10, nardini12, parker18}. \citet{takahashi10} demonstrate that the soft excess emission shown in the \suzaku\ spectrum of Ton~S180 can be described as a disc blackbody-shaped model with a temperature of kT=$0.075$\,keV. \citet{nardini12} found that the broad band \xmm\ and \suzaku\ spectra of Ton~S180 can be modelled by a combination of two reflection components, one from the inner disc and one from a distant reflector. A more recent study by \citet{parker18} succssfully model the broad band \xmm\ spectrum of Ton~S180 with a combination of a soft Comptonisation component, a hard Comptonisation component from the corona, and a relativistic disc reflection component. The soft Compotonisation component accounts for the soft excess. However the relativistic reflection component requires a very high iron abundance $Z_{\rm Fe}>9Z_{\odot}$.

We first fit the stacked EPIC-pn spectrum of Ton~S180 with an  absorbed power-law model. The corresponding ratio plot is shown in Fig.\,\ref{pic_fe}. Similar to previous analysis, a broad emission line in the iron band and a strong soft excess below 3~keV are found in the spectrum. Second, we model the broad band spectrum with \texttt{MODEL2}. \texttt{MODEL2} offers a very good fit of both the broad emission line feature and the soft excess emission with C-stat/$\nu$=301.71/179. The best-fit model is shown in Fig.\,\ref{pic_final3} and the best-fit parameters can be found in Table A2. By modelling the broad line and the soft excess with the same model, we obtain a high BH spin of $a_*>0.98$. A very steep emissivity profile is found, indicating a very compact coronal region. A disc viewing angle of $36^{+4}_{-6}$$^{\circ}$ is found, which is consistent with the previous measurement \citep[e.g. $\approx 39^{\circ}$,][]{parker18}. As shown from the plot of the best-fit model, the spectrum is dominated by the disc reflection component with a reflection fraction $f_{\rm refl}=1.9\pm0.7$, making it the second highest reflection fraction in our sample. A modest high disc density of $\log(n_{\rm e})=15.6^{+0.3}_{-0.2}$ is required by spectral fitting.

\subsection{UGC~6728}

UGC~6728 is a Seyfert 1 galaxy \citep{veron06} that has one of the known lowest black hole masses \citep[ $7.1\times10^{5}M_{\odot}$,][]{bentz15}. Only one \xmm\ observation is available in the archive. The EPIC-pn observation of UGC~6728 is dominated by flaring particle background. Therefore, we use EPIC-MOS observations instead. The MOS spectra of UGC~6728 show tentative evidence for emission features in the iron band, similar with PG~1426$+$015, and a strong soft excess below 2~keV. 

We model the broad band spectrum with \texttt{MODEL2}. The relativistic reflection model accounts mainly for the soft excess. The BH spin parameter is not constrained so we assume a maximum BH spin during the fit. \texttt{MODEL2} offers a good fit with C-stat/$\nu$=254.61/209. The best-fit model is shown in Fig.\,\ref{pic_final3}. Only an upper limit of the disc density $\log(n_{\rm e})<18$ is found. The best-fit model predicts a very strong broad Fe~K emission line, which cannot be resolved by current data quality. More future observations with longer exposures will enable us to study the iron band of UGC~6728 in more details.

\begin{figure*}
\centering
\includegraphics[width=14cm]{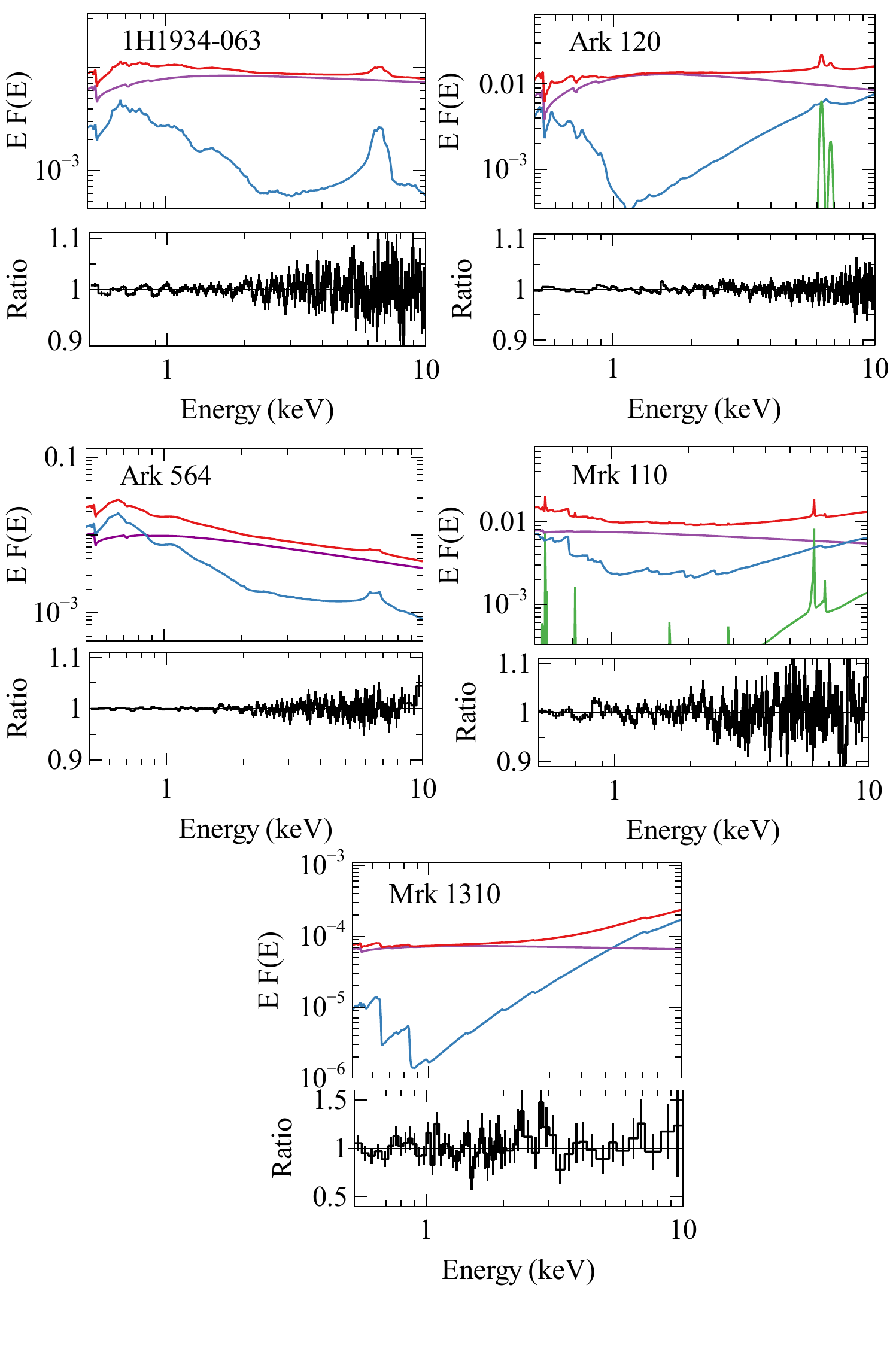}
\caption{The best-fit model and corresponding ratio plot for each source. Red: total model; blue: relativistic reflection model; green: distant reflector; purple: power-law shaped coronal emission. The unit of the y-axis in the model plots is keV\,cts\,cm$^{-2}$\,s$^{-1}$.}
\label{pic_final1}
\end{figure*}

\begin{figure*}
\centering
\includegraphics[width=14cm]{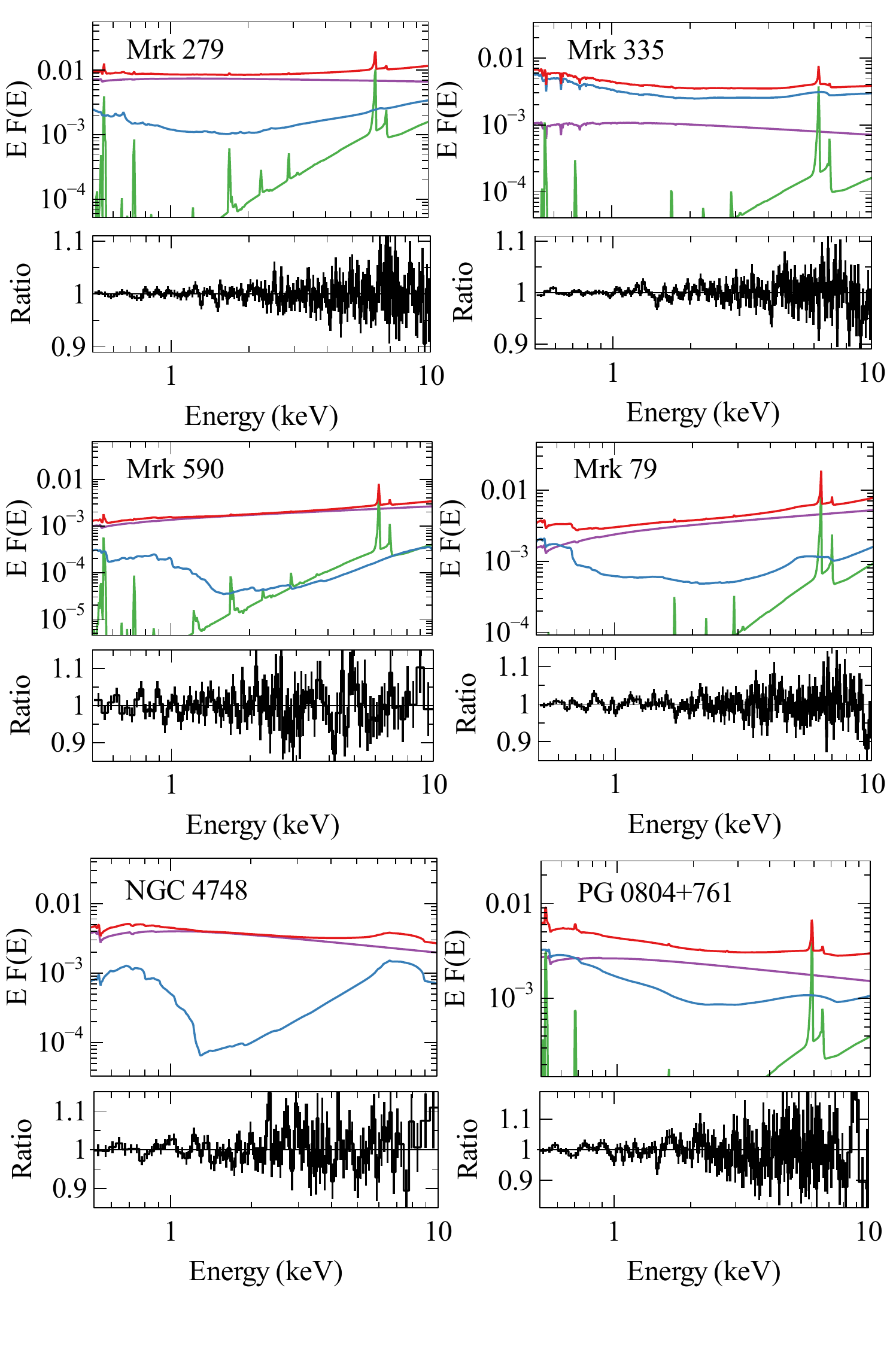}
\caption{Continued.}
\label{pic_final2}
\end{figure*}

\begin{figure*}
\centering
\includegraphics[width=14cm]{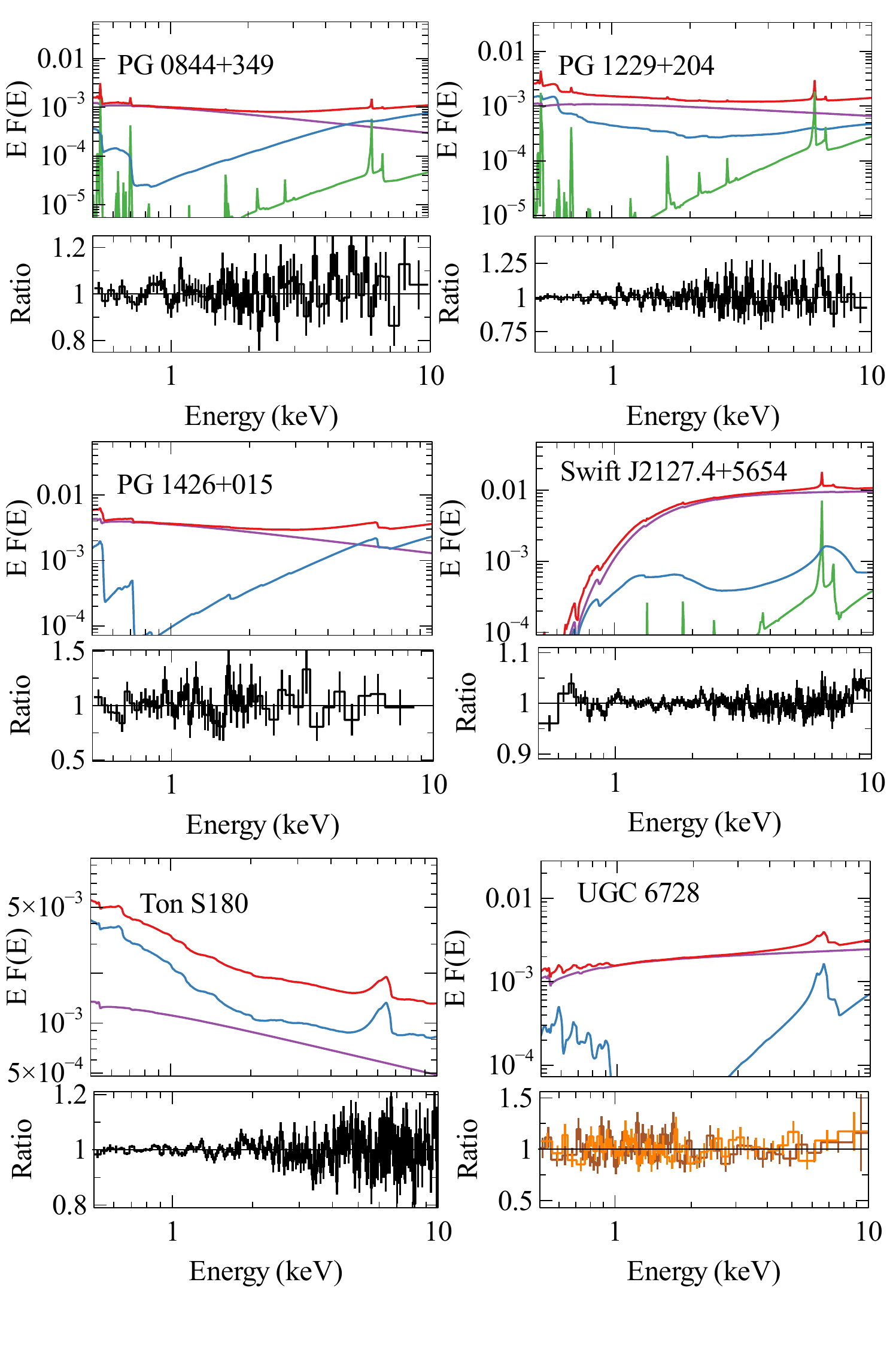}
\caption{Continued.}
\label{pic_final3}
\end{figure*}

\begin{table*}
\caption{A list of all the \xmm observations considered in this work. SW: small window mode; FF: full frame mode; LW: large window mode. Expo: Net exposure after correcting flaring particle background}
\label{tab_obs_info}
\centering
\begin{tabular}{cccccccccc}
\hline\hline
Name & obs ID & Expo (ks) & Date & pn mode & Name & obs ID & Expo (ks) & Date & pn mode \\
\hline
1H1934 & 0550451701 & 12 & 2009-04-28 & SW & Mrk~335 & 0101040101 & 28 & 2000-12-25 & FF\\
       & 0761870201 & 93 & 2015-10-03 & SW &         & 0306870101 & 90 & 2006-01-03 & SW\\
Ark~120 & 0147190101 & 76 & 2003-08-24 & SW&        & 0510010701 & 16 & 2007-07-10 & LW\\
        & 0693781501 & 85 & 2013-02-18 & SW&        & 0600540501 & 53 & 2009-06-13 & FF\\
        & 0721600201 & 86 & 2014-03-18 & SW&        & 0600540601 & 76 & 2009-06-11 & FF\\
        & 0721600301 & 85 & 2014-03-20 & SW&        & 0741280201 & 36 & 2015-12-30 & FF\\
        & 0721600401 & 85 & 2014-03-22 & SW&Mrk~590 & 0109130301 & 7 & 2002-01-01 & SW \\
        & 0721600501 & 87 & 2014-03-24 & SW&		& 0201020201 & 29 & 2004-07-04 & SW \\
Ark~564 & 0006810101 & 7.4& 2000-06-17 & SW&Mrk~79  & 0103862101 & 5.3 & 2000-10-09 & SW\\
        & 0006810301 & 7.0& 2001-06-19 & SW&		& 0400070201 & 14 & 2006-09-30 & SW\\
        & 0206400101 & 69& 2005-01-05 & SW&        & 0400070301 & 14 & 2006-09-30 & SW\\
        & 0670130201 & 41& 2011-05-24 & SW&        & 0400070401 & 14 & 2007-03-19 & SW\\
        & 0670130301 & 39& 2011-05-31 & SW&        & 0502091001 & 53 & 2008-04-26 & SW\\
        & 0670130401 & 38& 2011-06-06 & SW&NGC~4748 & 0723100401 & 26 & 2014-01-14 & LW \\
        & 0670130501 & 47& 2011-06-12 & SW&PG~0804 & 0605110101 & 16 & 2010-03-10 &  SW\\
        & 0670130601 & 42& 2011-06-17 & SW& & 0605110201 & 16 & 2010-03-12 & SW\\
        & 0670130701 & 32& 2011-06-26 & SW&PG~0844 & 0103660201 & 6 & 2000-11-04 & FF\\
        & 0670130801 & 29& 2011-06-29 & SW&              & 0554710101 & 12 & 2009-05-03 & LW\\
        & 0670130901 & 39& 2011-07-01 & SW&PG~1229 & 0301450201 & 17 & 2005-07-09 & SW\\
Mrk~110 & 0201130501 & 33 & 2004-11-15 & SW&PG~1426 & 0102040501 & 0.6 & 2000-07-28 & FF\\
Mrk~1310 & 0723100301 & 35 & 2013-12-09 & SW&UGC~6728 & 0312191601 & 8.4, 8.8 & 2006-02-23 &FF\\ 
Mrk~279 & 0083960101 & 13 & 2002-05-07 & FF&Ton~S180 & 0110890401 & 21 & 2000-12-14 &SW\\
        & 0302480401 & 40 & 2005-11-15 & SW& & 0110890701 & 13 & 2002-06-30 & SW\\
        & 0302480501 & 38 & 2005-11-17 & SW& & 0764170101 & 93 & 2015-07-03 & SW\\
        & 0302480604 & 22 & 2005-11-19 & SW& & 0790990101 & 21 & 2016-06-13 & SW\\
%			  & 0102040901 & 0 & 2000-07-28 & FF\\
 Swift~J2127 & 0601741901 & 5 & 2009-11-11 & FF \\
& 0655450101 & 84 & 2010-11-29 & SW\\
& 0693781701 & 94 & 2012-11-04 & SW\\
& 0693781801 & 93 & 2012-11-06 & SW\\
& 0693781901 & 50 & 2012-11-08 & SW\\
%1H0419$-$577& 0148000201   & 11.5  & 2002-09-25  & LW& Swift~J2127 & 0601741901 & 5 & 2009-11-11 & FF \\
%& 0148000301 & 0.3 & 2002-12-27      & LW && 0655450101 & 84 & 2010-11-29 & SW\\
%& 0148000401 & 11.0 & 2003-03-30     & LW  &                   & 0693781701 & 94 & 2012-11-04 & SW\\
%& 0148000501  & 5.8  & 2003-06-25    & LW  && 0693781801 & 93 & 2012-11-06 & SW\\
%& 0148000601  & 11.3  & 2003-09-16   & LW  &                   & 0693781901 & 50 & 2012-11-08 & SW\\
%& 0604720301 & 71.0 & 2010-05-30     & SW\\
%& 0604720401 & 42.3 & 2010-05-28     & SW   \\
\hline\hline
\end{tabular}
\end{table*}

\begin{landscape}
\begin{table}
\caption{Best-fit parameters for each source. The flux of each model component is calculated between 0.5--10~keV by using the convolution model \texttt{cflux} in XSPEC in the unit of erg\,cm$^{-2}$\,s$^{-1}$. The reflection fraction is defined as $F_{\rm refl}/F_{\rm pl}$ for simplicity.}
\label{tab_fit_info_high_ne}
\centering
\begin{tabular}{ccccccccccccccccccccc}
\hline\hline
Name & $q_{1}$ & $q_{2}$ & $R_{\rm r}$ & $a_{*}$ & $i$ (deg) & $\log(\xi)$ & $Z_{\rm Fe}$ & $\log(n_{\rm e})$ & $\Gamma$ & $\log(F_{\rm refl})$ & $\log(F_{\rm pl})$ & $\log(F_{\rm dis})$ & $f_{\rm refl}$ & C-stat/$\nu$ \\ 
\hline
1H1934 & $1.9^{+1.2}_{-0.3}$ & - & - & $>0.45$ & $44^{+4}_{-3}$ & $2.79\pm0.02$ & $5.9^{+0.6}_{-1.4}$ & $17.7^{+0.2}_{-0.3}$ & $2.106\pm0.005$ & $-11.02\pm0.03$ & $-10.387\pm0.004$ & - & $0.232^{+0.017}_{-0.016}$ & 242.35/180 \\
Ark~120 & $>8$ & $1.7^{+2.1}_{-0.2}$ & $<4$ & $>0.85$ & $67^{+4}_{-5}$ & $0.33^{+0.05}_{-0.04}$ & $0.7^{+1.0}_{-0.2}$& $<15.6$ & $2.284^{+0.010}_{-0.006}$ &$-10.745^{+0.007}_{-0.008}$ & $-10.1853^{+0.0019}_{-0.0007}$ & - & $0.276\pm0.005$ & 288.61/168 \\
Ark~564 & $5.62^{+0.8}_{-0.7}$ & $0.8^{+0.4}_{-0.3}$ & $14^{+7}_{-5}$ & $>0.9$ & $57^{+4}_{-3}$ & $2.723\pm0.003$ & $3.4^{+0.2}_{-0.4}$ & $18.55\pm0.07$ & $2.496^{+0.003}_{-0.002}$ & $-10.548^{+0.013}_{-0.008}$ & $-10.383^{+0.005}_{-0.009}$ & - & $0.68^{+0.03}_{-0.02}$ & 191.32/180\\
Mrk~110 & $8\pm2$ & $3\pm2$ & $4.3^{+1.2}_{-0.2}$ & $0.994^{+0.002}_{-0.008}$ & $35^{+3}_{-4}$ & $2.08^{+0.07}_{-0.31}$ & $0.7^{+0.5}_{-0.2}$ &  $<16.5$ & $2.15^{+0.04}_{-0.02}$ & $-10.74\pm0.02$ & $-10.480^{+0.004}_{-0.003}$ & $-11.75^{+0.08}_{-0.12}$ & $0.55\pm0.03$ & 205.35/161 \\
Mrk~1310 & 3 (f) & - & - & 0.998 (f) & $36^{+12}_{-8}$ & $<0.35$ & $<2.7$ & $17.0^{+0.4}_{-0.2}$ & $2.08^{+0.12}_{-0.10}$ & $-12.75^{+0.11}_{-0.10}$ & $-12.46\pm0.03$ & - & $0.51^{+0.17}_{-0.11}$ & 109.90/94  \\
Mrk~279 & $>8.5$ & $4^{+0.5}_{-1.2}$ & $3.1^{+1.2}_{-0.2}$ & $>0.95$ & $37^{+8}_{-17}$ & $2.29^{+0.19}_{-0.41}$ & $0.9\pm0.2$ & $<16.9$ & $2.064^{+0.038}_{-0.017}$ & $-11.10^{+0.02}_{-0.03}$ & $-10.450^{+0.008}_{-0.017}$ & $-11.64^{+0.06}_{-0.07}$ & $0.225^{+0.013}_{-0.017}$ & 155.59/177 \\
Mrk~335 & $8.0^{+1.8}_{-0.4}$ & $3.5^{+0.2}_{-0.4}$ & $2.80^{+0.20}_{-0.12}$ & $>0.988$ &  $33^{+4}_{-5}$ & $3.00^{+0.02}_{-0.04}$ & $2.3^{+0.4}_{-0.6}$ & $<16.0$ & $2.225\pm0.002$ & $-10.78\pm0.02$ &  $-11.30^{+0.10}_{-0.08}$ & $-12.40^{+0.03}_{-0.02}$ & $3.3^{+0.4}_{-0.6}$ & 251.39/169 \\
Mrk~590 & $6^{+3}_{-4}$ & $2\pm2$ & $4^{+7}_{-2}$ & $>0.1$ & $79^{+7}_{-4}$ & $1.4^{+0.5}_{-0.3}$ & $1.7^{+1.5}_{-1.2}$ & $18.4^{+0.2}_{-1.1}$ & $1.74^{+0.08}_{-0.07}$ &$-12.11^{+0.10}_{-0.12}$ & $-11.050^{+0.005}_{-0.010}$ & $-12.31\pm0.10$ & $0.08\pm0.02$ & 187.41/174 \\
Mrk~79 & $7.8^{+2.1}_{-2.3}$ & $3.0^{+0.6}_{-0.4}$ & $9^{+12}_{-6}$ & $>0.5$ & $21\pm6$ & $2.5^{+0.3}_{-0.2}$ & $1.6\pm0.5$ & $18.01^{+0.12}_{-0.32}$ & $1.712^{+0.014}_{-0.012}$  & $-11.32\pm0.03$ & $-11.93\pm0.07$ & $-10.775\pm0.012$ & $0.28\pm0.02$ & $225.60/177$  \\
NGC~4748 & $>5$ & $2\pm2$ & $<4.2$ & $>0.8$ & $62^{+2}_{-6}$ & $1.0^{+0.2}_{-0.3}$ & $3.9^{+2.4}_{-0.3}$ & $16.6^{+0.4}_{-0.5}$ & $2.38^{+0.02}_{-0.03}$ & $-11.48\pm0.02$ & $-10.750^{+0.012}_{-0.013}$ & - & $0.186^{+0.011}_{-0.010}$ & 152.66/144 \\
PG~0804 & $8.2^{+1.0}_{-2.3}$ & $3\pm2$ & $<4.2$ & $>0.94$ & $65^{+14}_{-9}$ & $2.98^{+0.04}_{-0.19}$ & $0.91^{+0.68}_{-0.09}$ & $<15.8$ & $2.26\pm0.03$ & $-11.06\pm0.02$ & $-11.0042^{+0.0012}_{-0.0008}$ & $-12.07^{+0.10}_{-0.11}$ & $0.88\pm0.04$ & 226.26/173 \\
PG~0844 & $3^{+3}_{-2}$ & $2\pm2$ & $<60$ & $>0.95$ & $32^{+12}_{-15}$ & $<0.06$ & $1.4^{+0.6}_{-0.2}$ & $17.02^{+0.06}_{-0.27}$ & $2.54^{+0.03}_{-0.05}$ & $-11.95^{+0.03}_{-0.02}$ & $-11.432^{+0.004}_{-0.006}$ & $-12.96^{+0.15}_{-0.12}$ & $0.306^{+0.023}_{-0.014}$ & 163.0/129 \\
PG~1229 & $>6$ & $4^{+1.5}_{-2.0}$ & $5^{+2}_{-3}$ & $0.93^{+0.06}_{-0.02}$ & $22^{+18}_{-10}$ & $2.3^{+0.7}_{-0.9}$ & $<1.5$ & $16.8^{+0.2}_{-0.3}$ & $2.24^{+0.13}_{-0.08}$ & $-11.71^{+0.08}_{-0.12}$ & $-11.34^{+0.03}_{-0.02}$ & $-12.4^{+0.2}_{-0.3}$ & $0.43\pm0.10$ & 111.08/128 \\
PG~1426 & 3 (f) & - & - & 0.998 (f) & 30 (f) & $3.0^{+0.4}_{-0.5}$ & $1.3^{+1.2}_{-0.3}$ & $<15.9$ & $2.03^{+0.11}_{-0.14}$ & $-11.39^{+0.10}_{-0.17}$ & $-10.87^{+0.03}_{-0.02}$ & - & $0.30^{+0.09}_{-0.10}$ & 53.43/52\\
Swift~J2127 & $>6$ & $2.2^{+0.4}_{-1.2}$ & $3^{+2}_{-1}$ & $0.72^{+0.14}_{-0.20}$ & $67^{+3}_{-2}$ & $2.32\pm0.02$ & $4.0^{+1.2}_{-2.3}$ & $>18.7$ & $1.953\pm0.008$ & $-11.03^{+0.02}_{-0.05}$ & $-10.365^{+0.003}_{-0.004}$ & $-11.94^{+0.07}_{-0.18}$ & $0.217^{+0.013}_{-0.022}$ & 246.21/182 \\
Ton~S180 & $>8$ & $2.5^{+0.6}_{-0.4}$ & $3.5^{+0.8}_{-0.6}$ & $>0.98$ & $36^{+4}_{-6}$ & $3.32^{+0.10}_{-0.12}$ & $3\pm2$ & $15.8^{+0.4}_{-0.2}$ & $2.377^{+0.015}_{-0.020}$ & $-11.09^{+0.08}_{-0.10}$ & $-11.36^{+0.14}_{-0.20}$ & - & $1.9\pm0.7$ & 301.71/179\\
UGC~6728 & $2.4^{+3.0}_{-0.2}$ & - & - & 0.998 (f) & $17^{+20}_{-8}$ & $2.7^{+0.2}_{-0.3}$ & $2.5^{+1.2}_{-0.8}$ & $<18$ & $1.86^{+0.08}_{-0.04}$ & $-11.87^{+0.08}_{-0.14}$ & $-11.013^{+0.012}_{-0.013}$ &-& $0.14^{+0.03}_{-0.04}$ & 254.61/209\\
\hline\hline
\end{tabular}
\end{table}
\end{landscape}

\begin{table}
    \centering
    \caption{Best-fit model parameters for Ton~S180 when the soft excess is modelled by a soft \texttt{cutoff} model and a fixed disc density $\log(n_{\rm e})=15$ is assumed. $E_{\rm cut}$ is in the unit of keV. The units of other parameters are the same as in Table\,\ref{tab_fit_info_high_ne}.}
    \begin{tabular}{ccc}
    \hline\hline
    Model & Parameter & Value \\
    \hline
    \texttt{cutoff} &  $\Gamma$ & $3.32^{+0.12}_{-0.13}$\\
                    &  $E_{\rm cut}$ & $11^{+32}_{-7}$\\
                    &  Norm & $1.8^{+0.4}_{-0.6}\times10^{-3}$ \\
    \texttt{powerlaw} & $\Gamma$ & $2.04^{+0.10}_{-0.13}$ \\
                      & $\log(F_{\rm pl})$ & $-11.34^{+0.16}_{-0.37}$\\
    \texttt{relxilld} & $q_{1}$ & >8\\
                      & $q_{2}$ & $2.6^{+0.4}_{-0.2}$\\
                      & $R_{\rm r}$ & $5^{+3}_{-2}$\\
                      & $i$ & $38\pm2$\\
                      & $a_*$ & $0^{+1}_{-1}$\\
                      & $Z_{\rm Fe}$ & >8 \\
                      & $\log(\xi)$ & $3.0\pm0.1$\\
                      & $\log(n_{\rm e})$ & 15 (fixed)\\
                      & $\log(F_{\rm refl})$ & $-11.49^{+0.10}_{-0.08}$\\
    C-stat/$\nu$ & & 294.72/178 \\
    \hline\hline
    \end{tabular}
    \label{tab_tons180_cutoff}
\end{table}

\section{The Disc Density of a Stable Gas Pressure-Dominated Disc} \label{gas}

In this section, we propose a possible explanation for the overestimation of the disc density in GX~339$-$4 given by the classical disc theory: the disc of GX~339$-$4 is gas dominated in the hard state while the Seyfert 1 AGN in our sample, which show a significantly higher accretion rate than GX~339$-$4 in the hard state, are in the radiation pressure-dominated regime in the \citet[][SS73]{shakura73} model. The disc density distribution across the height of a gas pressure-dominated disc might be the solution to this density problem in GX~339$-$4 observations. 

\begin{figure}
    \centering
    \includegraphics[width=\columnwidth]{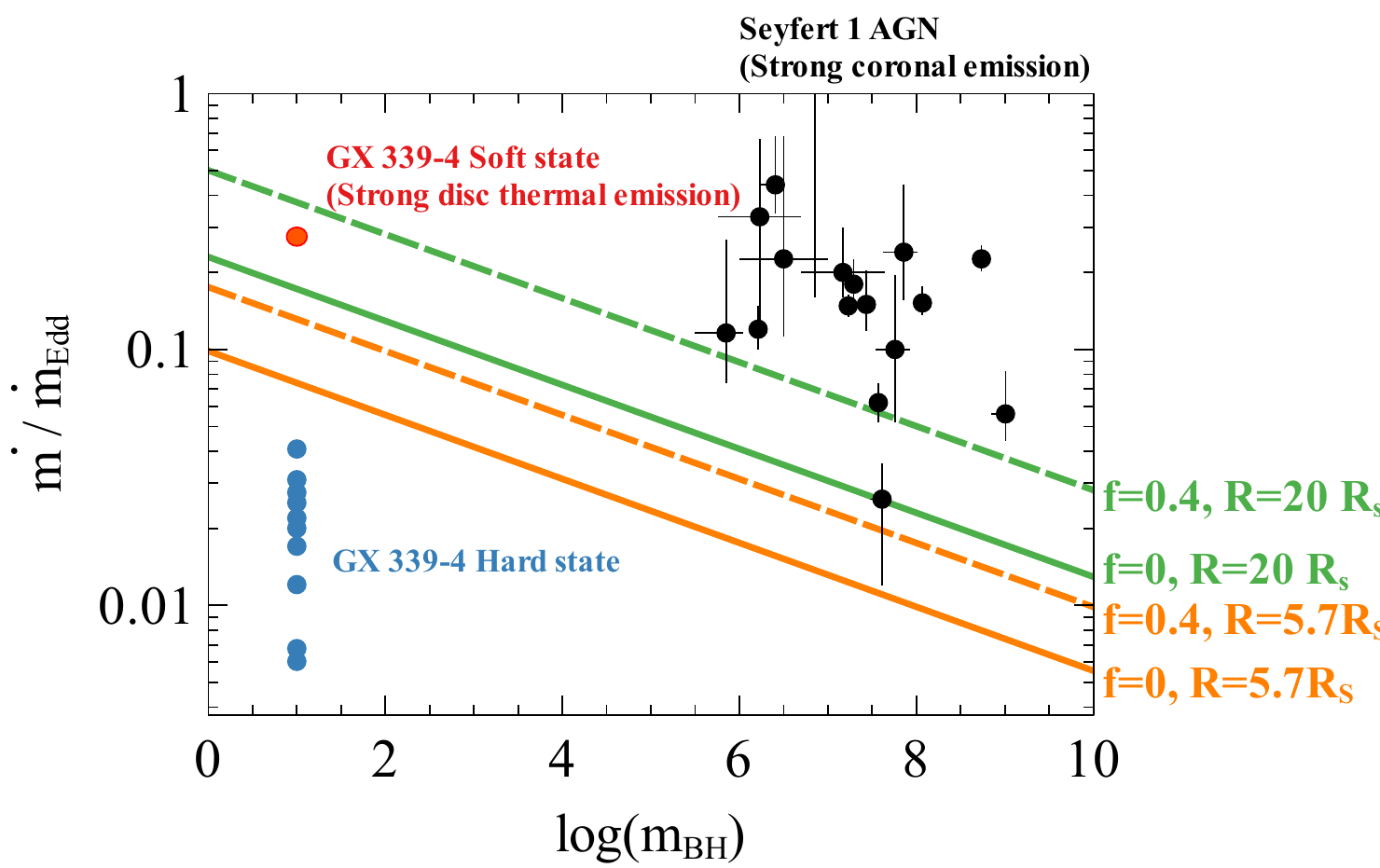}
    \caption{Mass accretion rates of 17 Seyfert 1 AGN and GX~339$-$4 that have been studied in our series. The mass accretion rates of GX~339$-$4 are calculated in our previous paper \citep{jiang19}. The orange lines show the threshold mass accretion rate below which the whole disc is dominated by gas pressure. The green lines show the threshold mass accretion rate below which $R<20\,R_{\rm S}$ is dominated by gas pressure. The solid lines are for $f=0$ is (SS73) and the dashed lines are for the scenario where 40\% of the disc energy is transferred to the coronal region \citep{svensson94}. }
    \label{pic_gas_disc}
\end{figure}

\subsection{Gas pressure-dominated disc}

In this section, we introduce the disc behaviour at different accretion rates in the SS73 model considering gas and radiation pressure in the disc. We assume that a fraction ($f$) of the disc energy is transferred to the coronal region instead of being radiated away locally from the disc \citep{svensson94}. The critical condition for the disappearance of the radiation pressure-dominated region in the disc is given by $P_{\rm gas}=P_{\rm rad}$, where $P_{\rm gas}$ and $P_{\rm rad}$ are respectively the gas and the radiation pressure of the disc. According to \citet{svensson94}, this condition at radius $r=R/R_{\rm S}=R/2r_{\rm g}$ leads to
\begin{equation}
    \frac{r}{(1-r^{-1/2})^{16/21}} = 188 (\alpha\ m_{\rm BH}/10^{8} )^{2/21} \dot{m}^{16/21} [\xi^{\prime}(1-f)]^{6/7},
    \label{eq_gas_radius}
\end{equation} 
$\alpha$ is the viscosity parameter and is assumed to be 0.1; $M_{8}$ is BH mass in the unit $10^{8}$M$_{\odot}$; $\dot{m}$ is defined as $\dot{m}=\dot{M}c^{2}/L_{\rm Edd}=L_{\rm Bol}/\epsilon L_{\rm Edd}$; $\xi^{\prime}$ is the conversion factor in  the radiative diffusion equation and assumed to be unity in the following calculation. When $f=0$, equation \ref{eq_gas_radius} reproduces the result in SS73.

The left side of equation \ref{eq_gas_radius} has a minimum value when $r\approx5.72$ \citep{svensson94}. For certain $m_{\rm BH}$ and $f$, the right side of equation \ref{eq_gas_radius} is always smaller than the left side when $\dot{m}$ is sufficiently small. In this case, there is no solution to equation \ref{eq_gas_radius} and the radiation pressure-dominated region in the disc disappears. The orange lines of Fig.\,\ref{pic_gas_disc} show this threshold mass accretion rate for $f=0$ and $f=0.4$. For sources that are below the orange lines, the whole disc is dominated by gas pressure in the classical disc theory. For larger $\dot m$, the radiation pressure-dominated region starts to expand from $r\approx5.72$ both inward and outward in the disc. For example, the green lines of Fig.\,\ref{pic_gas_disc} show the threshold mass accretion rate below which $r>20$ is dominated by gas pressure.    

The mass accretion rates of the 17 Seyfert 1 AGN in our sample and GX~339$-$4 are also shown in Fig.\,\ref{pic_gas_disc}. All Seyfery 1 AGN in our sample are above the green lines except Mrk~79. Mrk~79, which has a large uncertainty for the accretion rate estimation, is however still above the orange lines, indicating at least the inner region of the disc is dominated by radiation pressure. In comparison, the hard state of GX~339$-$4 shows a much lower accretion rate and the corresponding observations indicate the source is in the gas pressure-dominated regime. The soft state of GX~339$-$4 during an outburst however shows a much higher accretion rate and is the regime where the innermost region (e.g. $r\lessapprox20$) of the disc is dominated by radiation pressure. Note that the X-ray emission from GX~339$-$4 is dominated by a strong disc thermal component with a relatively weak power-law continuum during this soft state observation \citep{parker16,jiang19} while the Seyfert 1 AGN in our sample show very strong power-law emission in the X-ray band.

\subsection{Disc density in a gas pressure-dominated disc}

So far, we have shown the indication that the accretion disc of GX~339$-$4 in the hard state is in the gas pressure-dominated regime in the SS73 model. In this section, we calculate the disc density distribution across the height of a gas pressure-dominated disc in order to explain the density overestimation given by SS73. 

SS73 and \citet{svensson94} assume a thin disc with a uniform density across the height of a thin disc. The disc in reality however has a finite thickness which is particularly important for a gas pressure-dominated disc. Therefore, we calculate the surface disc density of a gas pressure-dominated disc assuming a simple iso-thermal disc model.

The standard thin disc model in SS73 gives the solutions of the number density $n_{\rm e}$ of a gas pressure-dominated disc:
\begin{equation}
n^{\rm gas}_{\rm e} = \frac{1}{\sigma_{\rm T} R_{\rm S}} K \alpha^{-7/10} r^{-33/20} \dot{m}^{2/5} [1-(r/r_{\rm in})^{-1/2}]^{2/5},
\label{eq_gas_ss73}
\end{equation}
where $K=2^{-7/2} (\frac{512\sqrt{2}\pi^3}{405})^{3/10}(\alpha_{\rm f}\frac{m_{\rm p}}{m_{\rm e}})^{9/10} (\frac{R_{\rm S}}{r_{\rm e}})^{3/10}$. $\alpha_{\rm f}$ is the fine-structure constant; $m_{\rm p}$ is the proton mass; $m_{\rm e}$ is the electron mass; $r_{\rm e}$ is the classical electron radius; $\sigma_{\rm T}=6.64\times10^{25}$cm$^{2}$ is the Thomson cross section. This solution is shown by the red dashed line in Fig.\,\ref{pic_ne2} for $m_{\rm BH}=10$. The inferred disc densities in the corresponding observations are 100 times lower than the predictions in SS73 (see Fig.\,\ref{pic_ne2}). 

%One of the possibilities is that SS73 considers a constant disc density in the vertical direction of a stable accretion disc while the reflection spectrum comes from an optical depth of $\tau<1$. 

Here we estimate the disc density distribution in the disc that is regulated by gas pressure based on the following assumptions: 1) the total luminosity of the accretion process depends on the accretion rate at the mid-plane of the disc; 2) the solution of SS73 (Eq.\,\ref{eq_gas_ss73}) applies only to the mid-plane of the disc. Then we calculate the disc density within an optical depth of $\tau=1$, where the observed reflection spectrum comes from.

We ignore the self gravitation of the disc. In the vertical direction of the disc, the pressure gradient is balanced by the gravitational force in the $z$ direction of the disc plane:
\begin{equation}
    \frac{dP}{dz}=-\rho \frac{GM}{R^{3}}z,
    \label{eq_rho_1}
\end{equation}
$P$ is the inner pressure of the disc; $M$ is the mass of the BH. 

For a gas-dominated disc, the inner pressure is
\begin{equation}
    dP=C^{2}_{\rm s}\ d\rho,
    \label{eq_rho_2}
\end{equation}
where $C_{\rm s}$ is the speed of sound and is about $6.3\times10^{4}$\,m\,s$^{-1}$ at $10^{7}$\,K for the disc of a BH XRB. 

By combining Eq.\,\ref{eq_rho_1} and Eq.\,\ref{eq_rho_2}, we obtain
\begin{equation}
    \frac{d\rho}{\rho}=-\frac{GM}{R^{3}C^{2}_{\rm S}}z\ dz.
\end{equation}

By integrating two sides, we obtain the disc density distribution of a gas-dominated disc:
\begin{equation}
    \ln{\frac{n_{\rm e}}{n_{0}}}=\ln{\frac{\rho}{\rho_{0}}}={-\frac{GM}{2R^{3}C^{2}_{\rm S}}z^{2}},
    \label{eq_rho_3}
\end{equation}
where $n_{0}$ and $\rho_{0}$ are the number density and the density at the mid-plane of the disc. We assume $n_{0}\approx n^{\rm gas}_{\rm e}$ given in Eq.\,\ref{eq_gas_ss73} and a pure hydrogen disc. For a 10$M_{\odot}$ BH that is accreting at $\dot{m}=0.1$, the disc has a number density of $n_{0}=1.0\times10^{23}$\,cm$^{-3}$ and an equivalent density of $\rho_{0}=167$\,kg\,m$^{-3}$ according to SS73 (see the red dashed line in Fig.\,\ref{pic_ne2}).  We define $J\equiv\frac{GM}{2R^{3}C^{2}_{\rm S}}$ and then the number density is given by $n_{\rm e}$=$n_{0}e^{-Jz^{2}}$. This density distribution in the vertical direction is consistent with Eq.5.33 in \citet{frank02}. Assuming $M=10M_{\odot}$ and $R=2r_{\rm g}$ as in Fig.\,\ref{pic_ne2}, the value of $J$ is approximately $6.22\times10^{-3}$\,m$^{-2}$.

Now we calculate the depth into the disc that is equivalent to the Thomson optical depth $\tau=1$. The optical depth is measured downwards from $+\infty$ to $z_0$,
\begin{equation}
    \tau=\int^{z_{0}}_{+\infty} - k \rho\ dz =k \rho_0 \int^{+\infty}_{z_{0}} e^{-Jz^{2}}\ dz = \frac{k \rho_0 }{\sqrt{J}} \int^{+\infty}_{\sqrt{J}z_0} e^{-x^{2}}\ dx,
    \label{eq_tau_1}
\end{equation}
where $k$ is the opacity of pure hydrogen for free-electron scattering with the same efficiency at all wavelengths and equals 0.04\,m$^{2}$\,kg$^{-1}$. The right side of Eq.\,\ref{eq_tau_1} should be equal to 1. By solving the Gaussian integral in Eq.\,\ref{eq_tau_1}, we obtain $\sqrt{J}z_{0}=1.76$ ($z_{0}=22$\,m). The density at $z_{0}$ is thus $4.5\%$ of the density in the mid-plane of the disc at $R=2r_{\rm g}$. Similarly, we consider $R=5r_{\rm g}$ and obtain $\sqrt{J}z_{0}=2.07$ and $n_{\rm e}=1.3\%n_0$. These fractions match the difference between our measurements and the solutions in SS73.

%%%%%%%%%%%%%%%%%%%%%%%%%%%%%%%%%%%%%%%%%%%%%%%%%%

% Don't change these lines
\bsp	% typesetting comment
\label{lastpage}
\end{document}